\documentclass[12pt,draftcls,onecolumn]{IEEEtran}
\usepackage{CJK}
\usepackage{amsmath}
\usepackage{color,latexsym,amsfonts,amssymb}
\usepackage{graphicx}      
\usepackage{graphicx}      
\usepackage{epsfig} 
\usepackage{mathptmx} 
\usepackage{times}
 \newtheorem{theorem}{Theorem}

\newtheorem{assumption}{Assumption}

\ifCLASSINFOpdf
\else
\fi
\begin{document}

\def\ba{\begin{array}}
\def\ea{\end{array}}
\def\ban{\begin{eqnarray*}}
\def\ean{\end{eqnarray*}}
\def\bd{\begin{description}}
\def\ed{\end{description}}
\def\be{\begin{equation}}
\def\ee{\end{equation}}
\def\bna{\begin{eqnarray}}
\def\ena{\end{eqnarray}}
%
\title{Distributed Economic Dispatch for Energy Internet Based
on Multi-Agent Consensus Control}
%
%
%

\author{Wushun~Chen  and
        Tao~Li,~\IEEEmembership{Senior Member,~IEEE}
\thanks{*Corresponding author: Tao Li. This work was supported in part by the National Natural Science Foundation of China under
grant 61522310 and the Shu Guang project of Shanghai Municipal Education Commission and Shanghai
Education Development Foundation under grant 17SG26. Partial results were presented at the 7th
IFAC Workshop on Distributed Estimation and Control in
Networked Systems (NecSys 2018),
Groningen, the Netherlands, August 27-28, 2018.}
\thanks{W. Chen is with School of Mechatronic Engineering and Automation, Shanghai University, Shanghai~200444, China (e-mail: cccws@qq.com).}
\thanks{T. Li is with the Shanghai Key Laboratory of Pure Mathematics and Mathematical Practice, School of Mathematical Sciences, East China Normal University, Shanghai 200241, China (e-mail: tli@math.ecnu.edu.cn).}
}

%
%

\markboth{Journal of \LaTeX\ Class Files,~Vol.~14, No.~8, August~2015}%
{Shell \MakeLowercase{\textit{et al.}}: Bare Demo of IEEEtran.cls for IEEE Journals}
%



\maketitle

\begin{abstract}
We consider the economic dispatch (ED) for an Energy Internet composed of energy routers (ERs), interconnected microgrids and main grid. The
microgrid consists of several bus nodes associated with distributed generators (DGs) and
intelligent control units (ICUs).
We  propose a distributed ED algorithm for the grid-connected microgrid, where
each ICU iterates the estimated electricity price of the distribution system and the estimation for the average power mismatch of
the whole microgrid by leader-following and average consensus algorithms, respectively. The ER iterates the incremental power exchanged with the distribution system.
By constructing an auxiliary consensus system, we prove that if the communication topology of the Energy Internet contains a spanning tree with the ER as the root and there is a path from each ICU to the ER, then
the estimated electricity price of the distribution system converges to its real value,
the power supply and demand achieves
balance and the ED achieves optimal asymptotically. Furthermore, we propose an autonomous distributed ED algorithm covering both grid-connected and isolated modes of the microgrid by feeding back the estimated average power mismatch for updating the incremental costs with penalty factor.
It is proved that if the communication topology of the microgrid is connected and there exists an ICU bi-directionally neighboring the ER, then the microgrid can switches between the two modes reliably. The simulation results demonstrate the effectiveness of the proposed algorithms.

\end{abstract}


%
\IEEEpeerreviewmaketitle

\section{Introduction}
\begin{figure}[!t]
\centering
\includegraphics[width=4in]{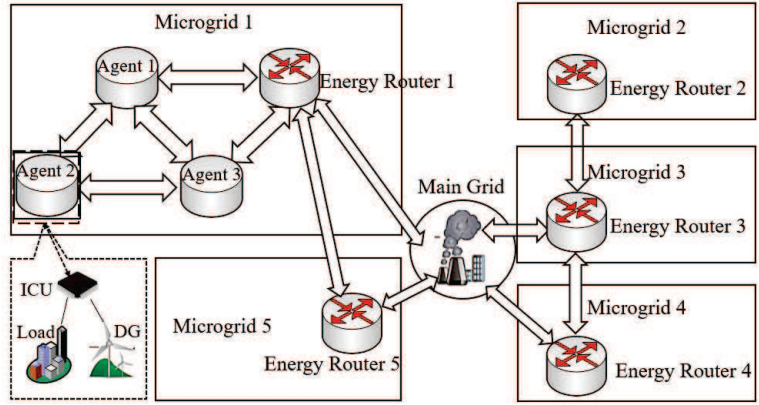}
\caption{An Energy Internet based on multi-agent systems.}
\label{fig1}
\end{figure}

%
%
%
%
\IEEEPARstart{R}{enewable}
power generation technologies, such as wind and
solar power generation, are promoted and used more and more
widely, which can relieve the shortage of fossil energy and avoid  environmental pollution. However, the characteristics of these renewable energy generations such as intermittency and uncertainty pose great challenges to the control
and optimization of power systems (\cite{lopes2007integrating}-\cite{awad2008control}). For
distributed generation of renewable energy sources, microgrids
are really flexible and efficient. A microgrid is composed of
distributed generators (DGs), energy storage devices, loads, and intelligent control units (ICUs), which is widely used for the grid planning
and optimization control of the integration of numerous and
diverse renewable energy generators. In recent
years, along with the rapid development of ``Internet+'' industries and Cyber-Physical systems, the concept of Energy Internet
has emerged. An Energy Internet is a combination of internet, renewable energy generation and
smart grid technologies,
which is essentially a Cyber-Physical energy system (\cite{Zhou2017development}-\cite{Liu2014Dynamic}). In an Energy Internet, the main power grid is
the ``backbone network'', microgrids are local area networks,
and energy routers (ERs) are intermediate ICUs among microgrids and external networks. This ultimately realizes the distributed and autonomous
cooperative management of power systems by a bottom-up structure (\cite{huang2011future}-\cite{cao2014energy}). In an Energy Internet, the local loads and DGs are directly controlled by local controllers, which
are called ICUs, equipped on each bus node of the microgrid system. ICUs can exchange their state information with other ICUs of neighboring bus nodes. As intermediate
units connecting microgrids and the external network,
ERs play roles in interconnecting each microgrid to
the distribution system,
and meeting the balance of power  supply and demand
of microgrids through power exchange.
The main grid  plays a role in broadcasting the electricity price to
microgrids and exchange power with the microgrids when the power supply
and demand is unbalanced among them. In
an Energy Internet, ICUs, ERs,
microgrids, and the main grid, can be viewed as agents on different levels. The architecture of an Energy
Internet based on multi-agent systems is shown in Fig. \ref{fig1}.

Economic dispatch problem (EDP) is an active research direction of power systems (\cite{Lin1984Hierarchical}-\cite{Chaturvedi2009Particle}). For EDPs, it is studied how to minimize the total generation cost by reasonably assigning the active power of each generator subjected to the balance of power supply and demand and generation limits. Many kinds of centralized ED algorithms have been investigated (\cite{Lin1984Hierarchical}-\cite{Chaturvedi2009Particle}). In all centralized algorithms, a central controller is needed with the knowledge of total states and
parameters of all bus nodes of the microgrid system. This requires a
very powerful communication infrastructure. In addition,  if the central controller is under attack,
then the whole microgrid system will break down. For an
Energy Internet, it is clear that distributed energy management algorithms are fundamental for restricting the complexity of
controller synthesis with the size of the system and are more
suitable than centralized algorithms for the flexibility and scalability of the grid topology and the plug-and-play feature of
DGs and loads in microgrids.

As the most basic algorithms of distributed cooperation, multi-agent consensus control algorithms have been studied widely. According to whether there are external interveners (leaders), they can be divided into leader-following and leader-free algorithms, which both implement some kind of distributed estimation through information interaction among adjacent nodes. The leader-following algorithm guarantees that the state of each follower node tends to the state of the leader node, thereby, achieves the distributed estimation of the leader's state (\cite{Liu2008Controllability}). A typical leader-free algorithm is the average-consensus algorithm, so that for any initial states ${x_i(0), i = 1, 2..., N}$, the state of each node $x_i(t)$ tends to $\frac{1}{N}\sum_{j=1}^{N}x_j(0)$ (\cite{Olfati2004Consensus}-\cite{Kingston2006Discrete}), thereby, achieves the distributed estimation of $\frac{1}{N}\sum_{j=1}^{N}x_j(0)$.
 Multi-agent consensus control algorithms have been used in EDPs for microgrids.
Zhang and Chow \cite{zhang2012convergence} propose a distributed ED method based on incremental cost consensus with a quadratic model of power generation costs under an undirected communication topology. Binetti \emph{et al}. \cite{binetti2014distributed} study distributed EDP with transmission losses. The concept of time stamp is introduced in the estimation of total power mismatch, and after a finite number of communications, all nodes obtain the relatively up-to-date total power mismatch. Zhang \emph{et al}. \cite{zhang2011decentralizing} propose a two-level consensus algorithm. On the high level, the incremental cost of
each DG achieves consensus, and on the low level, the average power mismatch of all buses of the microgrid is iteratively estimated by the average-consensus algorithm, and
the limit value is used as feedback to update the incremental cost of each DG. Kar and Hug \cite{kar2012distributed} propose a ``consensus + innovation'' type algorithm  to ensure the balance of power supply and demand of the total system. Based on the algorithm proposed in \cite{zhang2011decentralizing}, in \cite{yang2012consensus}, the central node is removed,  the average power mismatch
of all buses of the  microgrid is iteratively estimated
by the average-consensus algorithm, and the limit value is used as feedback to update the incremental cost of each DG. Li \emph{et al}. \cite{Li2019Consensus} propose a ED algorithm combining frequency control and consensus algorithms under the assumption that the measured frequency is the same for all nodes.
Yang \emph{et al}. \cite{yang2016minimum} propose a minimum-time
consensus-based approach for ED of microgrids. Besides,   ED algorithms with uncertainties such as communication delays,  noises, packet dropouts and random switching of network topologies in real communication networks are studied in \cite{zhang2016robust}-\cite{xu2017adistributed}.

The above research mainly focuses on the case of a single isolated microgrid. For the case of multiple interconnected microgrids, Wu and Guan \cite{Wu2013Coordinated} propose a decentralized Markov decision process to simulate EDP of multiple interconnected microgrids, which minimizes the total operation cost. Huang \emph{et al}. \cite{huang2016distributed} propose two consensus algorithms, one of which drives the incremental cost of each DG to the electricity price of the main grid, and the other one is to estimate the active power supplied by the main grid. The algorithm takes an important step in the field of ED for Energy Internet, and realizes ED for the grid-connected operation mode. However, there is severe fluctuation of the active power supplied by the distribution system due to the one-off estimation of the total power mismatch of the microgrid, which restricts the practical application of the algorithm. Wang \emph{et al}. \cite{Wang2016Decentralized} propose a hierarchical two-layer algorithm for EDPs of a single microgrid and interconnected  multi-microgrid systems.

In this paper, we study EDP of an Energy Internet based on multi-agent systems. Different from \cite{zhang2012convergence, zhang2011decentralizing} and \cite{yang2012consensus}, the microgrid consists of a number of bus nodes with DGs, loads and ICUs, and is connected to the distribution system (the main grid and other microgrids) by the ER. The topology of the whole network is a digraph. We propose a distributed ED algorithm based on multi-agent consensus control and incremental power exchanging by the ER. Firstly, we consider the grid-connected case and all ICUs know that the microgrid is in the grid-connected mode. On one hand, each ICU iteratively estimates the electricity price of the distribution system obtained by the ER by a leader-following consensus algorithm. On the other hand, the average power mismatch of the whole microgrid is iteratively estimated by average consensus algorithm.
 During each iteration, the ER calculates the incremental active power exchanged with the distribution system for the next time in a distributed way. Compared with the one-off estimation of the total power mismatch in \cite{huang2016distributed}, our algorithm can reduce the fluctuation of the exchanged power with guaranteed convergence. This is more conducive to practical application.  At the  stage of incremental power exchanging with the distribution system, there is a coupling between the estimation of the average power mismatch of all bus nodes and the calculation of incremental power exchanged with the distribution system, which leads to difficulties for the convergence analysis of the algorithm. To this end, we develop a set of analytical methods combining algebraic graph theory, difference equation stability and limit theory. By constructing an auxiliary system, the asymptotic stability of the algorithm for estimating the average power mismatch is converted into the convergence of consensus algorithm with all the neighbor nodes of the ER being a virtual leader as a whole. It is proved that if the communication topology of the Energy Internet contains a spanning tree with the ER as the root, all the ICUs of the microgrid form an undirected graph and there is a path from each ICU to the ER, then the estimated electricity price of the distribution system converges to its real value, so that the whole microgrid system achieves the balance of power supply and demand and optimal ED asymptotically. Numerical simulations demonstrate the effectiveness of the proposed algorithm.

For an Energy Internet, microgrids usually have two operation modes, namely, isolated mode (island operation) and networked mode (grid-connected operation). The mode of a microgrid is usually determined by the ER. The ICUs in the microgrid should be autonomous and those who are not neighbors of the ER do not need to know the operation mode of the whole microgrid. Therefore, a good distributed ED algorithm should ensure the transparency of operation mode information of the microgrid to the internal ICUs, that is, even if the ICUs which are not neighbors of the ER do not know the operation mode of the whole microgrid, the smooth switching between isolated and grid-connected modes can be achieved. The distributed ED algorithm of a single microgrid is considered in  \cite{zhang2012convergence}-\cite{yang2016minimum}. The case with interconnected multiple microgrids are considered in \cite{Wu2013Coordinated}-\cite{Wang2016Decentralized}. Most of the above algorithms only cover a special operation mode of a given microgrid, and it is impossible for them to integrate both isolated and grid-connected modes together with a smooth transition. Motivated by the above considerations, we further propose a distributed ED algorithm which can switch between the two operation modes smoothly.  Based on the grid-connected algorithm, the estimated power mismatch is used as feedback to update the incremental costs with penalty factor, and at the stage of estimating the average power mismatch of all bus nodes of the microgrid, the power mismatch compensation mechanism is introduced so that the total power mismatch value is kept before and after the isolated/connected mode transition. The algorithm is fully distributed in the sense that each ICU operates only based on its own state information and those obtained from neighbors. The ICUs which are not neighbors of the ER do not need to know the operation mode of the whole microgrid. We prove that if the communication topology of the Energy Internet contains a spanning tree with the ER as the root, the communication topology of the microgrid is connected and there is at least one ICU neighboring the ER bidirectionally, then the microgrid can switch between the isolated and grid-connected modes reliably. Numerical simulations demonstrate the effectiveness of the proposed algorithm.

The remainder of this paper is organized as follows. The preliminary knowledge on mathematical models of EDP and graph theory is introduced in Section II. The distributed ED algorithm for the grid-connected mode is proposed in Section III. Furthermore, a distributed ED algorithm which can perform smooth switching between isolated and grid-connected modes is proposed in Section IV. The feasibility of the algorithms by simulation is demonstrated in Section V. Finally, the paper is summarized and some future research topics are given in Section VI.

\emph{Notation}: $\mathbb{R}^{n}$ denotes the $n$-dimensional Euclidean  space; $ {I_{n}}$ denotes the $n$-dimensional identity matrix; $ {0_{m\times n}}$ denotes the $m\times n$-dimensional zero matrix; $X^T$ denotes the transpose of a given vector or matrix $X$; $\| {X}\|$ represents the 2-norm of $ {X}$; Denote $ {J_N}=\frac{1}{N} {1_N1_N^T}$, where $ {1_N}$ denotes the $N$-dimensional vector whose elements are all $1$.

\section{PROBLEM FORMULATION AND PRELIMINARIES}

\subsection{Graph Theory}
Let $\mathcal{G}=\{\mathcal{V}, \mathcal{E}_{\mathcal{G}}, \mathcal{A}_{\mathcal{G}}\}$ be a weighted digraph, where $\mathcal{V}=\{1,2,...,N\}$ is the  node set, $\mathcal{E}_{\mathcal{G}}$ is the edge set, and each edge in $\mathcal{G}$ is represented by an ordered pair $(j,i)$. The edge $(j,i)\in\mathcal{E}_{\mathcal{G}}$ if and only if node $j$ can send information to node $i$ directly, then node $j$ is called the parent node of node $i$, and node $i$ is called the child node of node $j$. The set of all parent nodes of node $i$  is denoted by $N_i=\{j\in \mathcal{V}|(j,i)\in\mathcal{E}_{\mathcal{G}}\}$. The matrix $\mathcal{A}_{\mathcal{G}}=[a_{ij}]$ $\in$ $\mathbb{R}^{N\times N}$ is called the weighted adjacency matrix of the digraph $\mathcal{G}$. For any $i$, $j\in \mathcal{V}$, $a_{ij}\geq0$, and $a_{ij}>0$ $\Leftrightarrow$ $j\in N_{i}$. The matrix $L_{\mathcal{G}}=\mathcal{D}_{\mathcal{G}}-\mathcal{A}_{\mathcal{G}}$ is called the Laplacian matrix of $\mathcal{G}$, where $\mathcal{D}_{\mathcal{G}}=diag(deg_{in}(1),...,deg_{in}(N))$.

If $\mathcal{A}_{\mathcal{G}}$ is symmetric, then $\mathcal{G}$ is called an undirected graph. The digraph $\mathcal{G}$ is said to be strongly connected if there exists a path between any pair of nodes. A directed tree is a special digraph. It has only one node which has no parents but only children (called the root node), and each of other nodes has only one parent.  A
spanning tree of $\mathcal{G}$ is a directed tree whose node set is $\mathcal{V}$ and whose edge set is a subset of $\mathcal{E}_{\mathcal{G}}$.

\vskip 0.2cm

\newtheorem{lemm}{Lemma}
\begin{lemm}
\label{oaddhxxx}
Let $\overline{\mathcal{G}}_0$$=$$\{\{0,1,2,...,N\}$, $\mathcal{E}_{\overline{\mathcal{G}}_0}$, $\mathcal{A}_{\overline{\mathcal{G}}_0}\}$ be a directed graph and $\mathcal{G}_0=\{\{1,2,...,N\}, \mathcal{E}_{\mathcal{G}_0},\mathcal{A}_{\mathcal{G }_0}\}$ be a subgraph of $\overline{\mathcal{G}}_0$ satisfying
\begin{equation*}
 {\mathcal{A}_{\overline{\mathcal{G}}_0}=\left(
\begin{array}{cc}
0_{1\times1} & [1^T_L,0^T_{N-L}]\\
0_{N\times1}   & \mathcal{A}_{\mathcal{G}_0}
\end{array}\right)},
\end{equation*}
where $L\in\{1,2,...,N\}$. Denote $ {C}_0=\left( \begin{smallmatrix}  {I}_L &  {0}_{L\times(N-L)} \\  {0}_{(N-L)\times L} &  {0}_{(N-L)\times (N-L)} \end{smallmatrix} \right)$. If for any node $i\in\{L+1,L+2,...,N\}$ of $\mathcal{G}_0$, there is node $j\in\{1,2,...,L\}$  such that there is a path from $j$ to $i$, then the eigenvalues of $( {I_N}- {C}_0)( {I_N}-\mu_0 {L_{\mathcal{G}_0}})$ are all inside the unit disk of the complex plane,
where $L_{\mathcal{G}_0}$ is the Laplacian matrix of $\mathcal{G}_0$, and $\mu_0\in\left(0,{1}/{\max_{i=1,2...,N}\sum_{j=1}^{N} a_{ij}}\right)$.
\end{lemm}

\vskip 0.2cm

\emph{Proof}:
Consider a discrete-time linear time-invariant system
\begin{equation}
\label{xfangchen}
 {X}(k+1)=( {I_N}- {C}_0)( {I_N}-\mu_0 {L_{\mathcal{G}_0}}) {X}(k),\ k=0,1,2,...,
\end{equation}
where $ {X}(k)=[x_1(k),x_2(k),...,x_N(k)]^T$. From (\ref{xfangchen}), it follows that for any $X(0)\in\mathbb{R}^{N}$, ${x}_i(k)=0,\ k=1,2,...,\ i=1,2...,L$. And
\bna
\label{eadd11xx}
&&\lefteqn{{x}_i(k+1) }\cr
&=&{x}_i(k)+\mu_0\Big[\sum_{j=L+1}^Na_{ij}({x}_j(k)-{x}_i(k))+\sum_{j=1}^{L}a_{ij}({x}_j(k)-{x}_i(k))\Big]\cr
&=&{x}_i(k)+\mu_0\Big[\sum_{j=L+1}^Na_{ij}({x}_j(k)-{x}_i(k))+\sum_{j=1}^{L}a_{ij}(0-{x}_i(k))\Big]\cr
&=&{x}_i(k)+\mu_0\Big[\sum_{j=L+1}^Na_{ij}({x}_j(k)-{x}_i(k))+\overline{b}_i^\prime(0-{x}_i(k))\Big]\cr
&&k=1,2,...,\ i=L+1,...,N,
\ena
where $\overline{b}_i^\prime=\sum_{j=1}^{L}a_{ij}$. Since  for any node $i\in\{L+1,L+2,...,N\}$ of $\mathcal{G}_0$, there is node $j\in\{1,2,...,L\}$ such that there is a path from $j$ to $i$, therefore,  (\ref{eadd11xx}) is indeed a leader-following consensus algorithm with the node set $\{1,2,...,L\}$ being a zero state virtual leader as a whole. The leader-following consensus algorithm is a special case of distributed consensus algorithms with digraphs (\cite{RenBeardBook}). Then from $\mu_0\in(0,\frac{1}{\max_{i=1,2...,N}\sum_{j=1}^{N} a_{ij}})$ (See Theorem 2.20 in \cite{RenBeardBook}), we get
\begin{equation*}
\lim_{k\to\infty}x_i(k)=0,\ i=1,2,...,N,\ \forall\ X(0)\in\mathbb{R}^{N}.
\end{equation*}

Noticing the arbitrariness of $X(0)$, we know that the eigenvalues of $( {I_N}- {C}_0)( {I_N}-\mu_0 {L_{\mathcal{G}_0}})$ are all inside the unit disk of the complex plane.
\hfill $\blacksquare$

\subsection{Economic Dispatch}
Suppose that there is an $N$-bus microgrid system connected to the distribution system. Each bus contains a DG and a load, and each DG is equipped with an ICU as the local controller. The generation cost function of the $i$th DG is given by
\begin{equation*}
F_i(P_i)=\frac{(P_i-\alpha_i)^2}{2\beta_i}+\gamma_i,\ i=1,2,...,N,
\end{equation*}
where $P_i$ is the active power generated by the $i$th DG, $\alpha_i\leq0$, $\beta_i>0$, $\gamma_i\leq0$ are the cost coefficients. The so called
EDP is to minimize the total generation
cost subjected to the balance of power supply and demand as well as generation limits of DGs, which is formulated as follows.
\begin{equation}
\label{main}
\begin{array}{c}
\underset{\{P_i,i=1,...,N;P_{MG}\}}{\min} \ \sum_{i=1}^N F_i(P_{i})+\lambda_0P_{MG},\\
{s.t.} \ \sum_{i=1}^N P_{i}+P_{MG} =\sum_{i=1}^N P_{Di}+P_L(P_1,...,P_N),\\
\underline{P}_{i}\le P_{i}\le \overline{P}_{i},
\end{array}
\end{equation}
where $P_{MG}$ is the power exchanged between the microgrid and the distribution system, and $\lambda_0$ is the electricity
price of the distribution system obtained by the ER. $\overline{P}_i\geq0$ and $\underline{P}_i\geq0$ are effective lower and upper power limits of the $i$th DG, respectively, dependent on its physical power limits and maximum ramping rate (\cite{Fan1998Real}).  If there is no generator but only a load at bus $i$, then $\overline{P}_i=\underline{P}_i=0$. $P_{Di}\geq0$ is the load at bus $i$. $P_L(P_1,...,P_N)=\sum_{i=1}^{N}P_{Li}(P_{i})$ represents the power transmission loss, where $P_{Li}(P_{i})=B_iP_{i}^2$ is the transmission loss caused by the $i$th DG
(\cite{Soliman2011Modern}-\cite{wood1996power}), and $B_i>0$ is the loss factor.

Noticing that $D=\{P_i,i=1,...,N;P_{MG}|\sum_{i=1}^N P_{i}+P_{MG} =\sum_{i=1}^N P_{Di}+P_L,\underline{P}_{i}\le P_{i}\le \overline{P}_{i}\}$ is a bounded and closed subset of $\mathbb{R}^{N+1}$ and the cost function of (\ref{main}) to be optimized is continuous on $D$, the optimization problem (\ref{main}) must have a global minimum. The Lagrange multiplier method can be used to solve the above EDP. For any feasible point $P_i$, define the active constraint sets by
\begin{equation*}
\begin{split}
\Omega(P_i)=\{i|P_i-\overline{P}_i=0\},\\
\Gamma(P_i)=\{i|P_i- \underline{P}_i=0\}.
\end{split}
\end{equation*}
Denote $\overline{\nu}=[\overline{\nu}_1,...,\overline{\nu}_N]^T$ and $\underline{\nu}=[\underline{\nu}_1,...,\underline{\nu}_N]^T$. Let the
Lagrangian function
\begin{equation*}
\begin{split}
  &L(P_1,...,P_N, P_{MG},\lambda,\overline{\nu},\underline{\nu})\\
  =&\sum_{i=1}^N F_i(P_{i})+\lambda_0P_{MG}+\lambda(\sum_{i=1}^N P_{Di}+P_L-\sum_{i=1}^N P_{i}-P_{MG})+\sum_{i=1}^N\overline{\nu}_i(P_i-\overline{P}_i)+\sum_{i=1}^N\underline{\nu}_i(\underline{P}_i-P_i),
\end{split}
\end{equation*}
where $\lambda$, $\overline{\nu}_i$, $\underline{\nu}_i$, $i=1,2,...,N$ are the Lagrangian multipliers for each DG, respectively. It is known from the KKT necessity condition (\cite{bertsekas1999nonlinear}) that if $\{P_i^*$, $i=1,...,N$; $P_{MG}^*\}$ is a local minimum point of (\ref{main}), then there is unique $\lambda^*$, $\overline{\nu}^*=[\overline{\nu}_1^*,\overline{\nu}_2^*,...,\overline{\nu}_N^*]$ and $\underline{\nu}^*=[\underline{\nu}_1^*,\underline{\nu}_2^*,...,\underline{\nu}_N^*]$, such that the following conditions hold.
\begin{eqnarray}       
\label{budengshizu}
\left\{                  
\begin{array}{l}     
     \nabla_{\{P_1,...,P_N, P_{MG}\}}L(P_1^*,...,P_N^*, P_{MG}^*,\lambda^*,\overline{\nu}^*,\underline{\nu}^*)=0,\\
     \overline{\nu}_i^*\ge0,\quad i=1,2,...,N,\\
     \underline{\nu}_i^*\ge0,\quad i=1,2,...,N,\\
     \overline{\nu}_i^*=0,\quad i\notin\Omega(P_i^*),\\
     \underline{\nu}_i^*=0,\quad i\notin\Gamma(P_i^*).
\end{array}           
\right.              
\end{eqnarray}
This gives
\begin{eqnarray*}       
\left\{                  
\begin{array}{lll}     
    \lambda^*=\frac{(P_{i}^*-\alpha_i)}{\beta_i({1-\partial P_L/\partial P_{i}^*})},
     \\
    \lambda^*=\lambda_0,
\end{array}           
\right.i\notin\Omega(P_i^*)\cup\Gamma(P_i^*),            
\end{eqnarray*}
where $1/(1-\partial P_L/\partial P_{i}^*)$ is the penalty factor of $i$th DG. Then the unique global optimal solution to  (\ref{main}) is given by
\begin{equation}
\begin{split}
\label{jizhongadd1}     
P_{i}^*
=\left\{                  
\begin{array}{lll}     
     \frac{\beta_i\lambda_0+\alpha_i}{1+2B_i\beta_i\lambda_0}, &\underline{P}_i\le \frac{\beta_i\lambda_0+\alpha_i}{1+2B_i\beta_i\lambda_0}\le\overline{P}_i, \\
     \overline{P}_{i},  &\frac{\beta_i\lambda_0+\alpha_i}{1+2B_i\beta_i\lambda_0}>\overline{P}_i,\\
     \underline{P}_{i},  &\frac{\beta_i\lambda_0+\alpha_i}{1+2B_i\beta_i\lambda_0} <\underline{P}_{i},
\end{array}           
\right.
\end{split}             
\end{equation}
and
\begin{equation}
\label{jizhongm}
P_{MG}^*=\sum_{i=1}^N P_{Di}+\sum_{i=1}^{N}B_i(P_{i}^{*})^2-\sum_{i=1}^N P_{i}^*.
\end{equation}

\vskip 0.2cm

\emph{Remark 1}: If the microgrid system is disconnected from the distribution system, then $P_{MG}=0$. For this case, the problem (\ref{main}) degenerates into EDP of an isolated microgrid.
Denote the optimal solution of (\ref{main}) with $P_{MG}=0$ by $\{P_{i}^{*\prime}$, $i=1,...,N$ $\}$, then from  (\ref{budengshizu}), we get
\begin{equation}
\begin{split}
\label{jizhongnlam}
\lambda^{*\prime}=\frac{(P_{i}^{*\prime}-\alpha_i)}{\beta_i({1-\partial P_L/\partial P_{i}^{*\prime}})},\ i\notin\Omega(P_{i}^{*\prime})\cup\Gamma(P_{i}^{*\prime}).
\end{split}
\end{equation}
If $i\in\Omega(P_{i}^{*\prime})\cup\Gamma(P_{i}^{*\prime})$, $\forall \ i=1,2,...,N$, then $P_{i}^{*\prime}=\overline{P}_{i}$ or $P_{i}^{*\prime}=\underline{P}_i$, $\forall$ $i=1,2,...,N$.
To avoid this trivial case, we always assume that there is $i\in\{1, 2,..., N\}$ such that $i\notin\Omega(P_i^{*\prime})\cup\Gamma(P_i^{*\prime})$.  Especially, this implies $\sum_{i=1}^N \underline{P}_{i}<\sum_{i=1}^N P_{Di}+P_L<\sum_{i=1}^N \overline{P}_{i}$, then it is avoided that there is no feasible solution for the isolated operation mode.

From  (\ref{main}),  we get  $\sum_{i=1}^N P_{i}^{*\prime}=\sum_{i=1}^N P_{Di}+P_L(P_1^{*\prime},...,P_N^{*\prime})$, which together with (\ref{jizhongnlam}) gives
\begin{equation}
\begin{split}
\sum_{i\notin\Omega(P_i^{*\prime})\cup\Gamma(P_i^{*\prime})}\frac{\beta_i\lambda^{*\prime}+\alpha_i}{1+2B_i\beta_i\lambda^{*\prime}}-\sum_{i\notin\Omega(P_i^{*\prime})\cup\Gamma(P_i^{*\prime})}B_i\left(\frac{\beta_i\lambda^{*\prime}+\alpha_i}{1+2B_i\beta_i\lambda^{*\prime}}\right)^2\\
=\sum_{i=1}^NP_{Di}
+\sum_{i\in\Omega(P_i^{*\prime})}B_i\overline{P}_{i}^2+\sum_{i\in\Gamma(P_i^{*\prime})}B_i\underline{P}_i^2-\sum_{i\in\Omega(P_i^{*\prime})}\overline{P}_{i}-\sum_{i\in\Gamma(P_i^{*\prime})}\underline{P}_i.
\end{split}
\end{equation}
This determines a unique $\lambda^{*\prime}$. Then the optimal ED solution is given by
\begin{equation}
\begin{split}
\label{jizhongn}   
P_{i}^{*\prime}=\left\{                  
\begin{array}{lll}     
     \frac{\beta_i\lambda^{*\prime}+\alpha_i}{1+2B_i\beta_i\lambda^{*\prime}}, &\underline{P}_i\le \frac{\beta_i\lambda^{*\prime}+\alpha_i}{1+2B_i\beta_i\lambda^{*\prime}}\le\overline{P}_i, \\
     \overline{P}_{i},  &\frac{\beta_i\lambda^{*\prime}+\alpha_i}{1+2B_i\beta_i\lambda^{*\prime}}>\overline{P}_i,\\
     \underline{P}_{i},  &\frac{\beta_i\lambda^{*\prime}+\alpha_i}{1+2B_i\beta_i\lambda^{*\prime}} <\underline{P}_{i}.
\end{array}           
\right.             
 \end{split}
\end{equation}
For an isolated microgrid, it can be proved that the optimal solution satisfies that the incremental costs with penalty factor of DGs are all equal, and the system satisfies the balance of power supply and demand  (\cite{wood1996power}).

\vskip 0.2cm

\emph{Remark 2}:
 It is a centralized algorithm to calculate the optimal solution $P_{i}^*$ (or $P_{i}^{*\prime}$) directly by (\ref{jizhongadd1})-(\ref{jizhongm}) (or (\ref{jizhongnlam})-(\ref{jizhongn})).  Then a central controller is required to collect the parameters  $\{\alpha_i,\beta_i,\gamma_i,B_i,\overline{P}_{i},\underline{P}_{i}\},i=1,2,...,N$ of all bus nodes. This requires a very strong communication infrastructure. In addition, the whole microgrid system will break down in case that the central controller is under attack.

\vskip 0.2cm

Let $g$ denote the operation mode of the microgrid, where $g=1$ represents grid-connected mode and $g=0$ represents isolated mode. Suppose that the $N$-bus microgrid system and its connected ER form a digraph denoted by $\overline{\mathcal{G}}$$=$$\{\{0,1,2,...,N\}$, $\mathcal{E}_{\overline{\mathcal{G}}}$, $\mathcal{A}_{\overline{\mathcal{G}}}\}$. The node ${0}$ represents the ER, which determines the operation mode of the microgrid, and the remaining $N$ nodes, which form an undirected graph denoted by $\mathcal{G}=\{\{1,2,...,N\}, \mathcal{E}_{\mathcal{G}},\mathcal{A}_{\mathcal{G}}\}$, represent the ICUs at every bus nodes of the microgrid system.  The graph $\mathcal{G}$ is a subgraph of digraph
$\overline{\mathcal{G}}$ with
\begin{equation*}
 {\mathcal{A}_{\overline{\mathcal{G}}}=\left(
\begin{array}{cc}
0_{1\times1} & 1_N^T\mathcal{A}_{0*}\\
\mathcal{A}_{*0}1_N   & \mathcal{A}_{\mathcal{G}}
\end{array}\right)},
\end{equation*}
Here, $\mathcal{A}_{*0}$$=$$diag(a_{10},$$a_{20},...,$$a_{N0})$ represents the weighted adjacency matrix between the ER (node $0$) and ICUs (the nodes of $\mathcal{G}$), $a_{i0}=1$$\Leftrightarrow$$0\in N_i$, and $a_{i0}=0$$\Leftrightarrow$$0\notin N_i$; $\mathcal{A}_{0*}$$=$$diag(a_{01}, $$a_{02},...,$$a_{0N})$, where $a_{0i}=1$$\Leftrightarrow $$i\in N_0$,  and $a_{0i}=0$$\Leftrightarrow $$i\notin N_0$. An ICU is also called an agent.

 We aim to design a distributed ED algorithm to achieve the global optimal solution of (\ref{main}), that is, each ICU solves the optimal EDP based on its own parameters $\{\alpha_i,\beta_i,\gamma_i,B_i,\overline{P}_{i},\underline{P}_{i}\}$, its own state $[\lambda_i(k), P_i(k)]$ and the information obtained from its neighboring ICUs.

\section{Distributed Economic Dispatch Algorithm in Grid-Connected Mode}

Firstly, we consider the case that the microgrid is always in the grid-connected mode which means that $g = 1$, and all ICUs know that the microgrid is in the grid-connected mode.

 The algorithm is divided into three parts. In the first part, a leader-following consensus algorithm is used for each agent.
 \begin{equation}
 \begin{split}
\label{itlambda1}
  \lambda_{i}(k+1)=\lambda_{i}(k)+\epsilon_i\Big[\sum_{j\in {N_i}}a_{ij}(\lambda_j(k)-\lambda_i(k)) + ga_{i0}(\lambda_0-\lambda_i(k))\Big],
  \end{split}
\end{equation}
which is to drive the incremental cost with penalty factor $\lambda_{i}(k)$ of each DG  to the electricity price $\lambda_0$ of the distribution system obtained by the ER, where $\epsilon_i>0$ is the step size of the algorithm, and $\lambda_i(0)$ is any given initial value.

In the second part, each agent calculates the active power at time $k$:
\begin{equation}
\begin{split}
\label{itpi1}    
P_{i}(k)=\Phi_i(\lambda_{i}(k))=\left\{                  
\begin{array}{lll}     
         \frac{\beta_i\lambda_i(k)+\alpha_i}{1+2B_i\beta_i\lambda_i(k)}, &\underline{P}_i\le \frac{\beta_i\lambda_i(k)+\alpha_i}{1+2B_i\beta_i\lambda_i(k)}\le\overline{P}_i, \\
     \overline{P}_{i},  &\frac{\beta_i\lambda_i(k)+\alpha_i}{1+2B_i\beta_i\lambda_i(k)}>\overline{P}_i,\\
     \underline{P}_{i},  &\frac{\beta_i\lambda_i(k)+\alpha_i}{1+2B_i\beta_i\lambda_i(k)}<\underline{P}_{i},
\end{array}           
\right.              
 \end{split}
\end{equation}
where $P_{i}(k)$ represents the active power generated by the $i$th  DG at time $k$. For (\ref{itpi1}), when $1+2B_i\beta_i\lambda_i(k)=0$,  it is stipulated that if $\beta_i\lambda_i(k)+\alpha_i>0$, then $P_{i}(k)=\overline{P}_{i}$, while  if $\beta_i\lambda_i(k)+\alpha_i<0$,  then $P_{i}(k)=\underline{P}_i$.

In the third part of the algorithm, each agent estimates the average power mismatch of all bus nodes of the microgrid system by average-consensus algorithm:
\begin{equation}
\label{itpm1}
\left\{
\begin{array}{ll}
  \begin{split}
    y_i(k+1)=\Delta \widehat{P}_i(k)+\mu\Big[\sum_{j\in {N_i}}a_{ij}(\Delta \widehat{P}_j(k)-\Delta \widehat{P}_i(k))\Big]+\Delta P_i(k+1)-\Delta P_i(k),
    \end{split}
    \\
    \Delta \widehat{P}_i(k+1)= (1-a_{0i})y_i(k+1).
\end{array}           
\right.            
\end{equation}
Here, $\Delta P_i(k)=P_{Di}+P_{Li}(k)-P_i(k)$
is the power mismatch of the $i$th bus at time $k$,  $P_{Li}(k)=B_iP_i^2(k)$ is the line loss due to the $i$th DG at time $k$, $\Delta \widehat{P}_i(k)$ is the local estimate of agent $i$ for the average power mismatch of all buses. The equation $\Delta \widehat{P}_i(k+1)= (1-a_{0i})y_i(k+1)$ means that for the neighboring agents of the ER, there is  direct power replenishment by the ER after each iteration, and so their estimates for the power mismatch are zeros.

And at each iteration, the incremental power exchanged with the distribution system is adjusted by the ER:
\begin{equation}
\label{pmguji}
 P_{MG}(k+1)=P_{MG}(k)+\sum_{i=1}^{N} a_{0i}y_i(k+1),
\end{equation}
where $P_{MG}(k)$ is the power exchanged with the distribution system through the ER at time $k$ and $\sum_{i=1}^{N} a_{0i}y_i(k+1)$ is the incremental power exchanged with the distribution system.
The initial values $\Delta\widehat{P}_i(0)$, $i=1,2,...,N$ and $P_{MG}(0)$ are chosen such that $\sum_{i=1}^{N}\Delta\widehat{P}_i(0)+P_{MG}(0)=\sum_{i=1}^{N}\Delta P_i(0)$, which is satisfied by letting
$\Delta\widehat{P}_i(0)=\Delta P_i(0)$, $i=1,2,...,N$ and $P_{MG}(0)=0$.

\vskip 0.2cm

\emph{Remark 3}:
As intermediate units connecting microgrids and the external
network, ERs play roles in interconnecting
each microgrid to the distribution system, monitoring and
control of energy quality, as well as information and communication security, etc.

As an intermediate unit between the microgrid and the external network, the ER is not only an information medium but  also a bridge for power exchange.
The equation (\ref{pmguji}) shows that the ER is an information medium.
For the neighbors ICUs of the ER, each ICU transmits its estimate of average power mismatch to the ER at each iteration.
The ER then calculates the power needed for exchanging with the distribution system for the microgrid according to (\ref{pmguji}).
Then the ER plays as an interchange of power, and the distribution system supplies (obtains) power to (from) the microgrid through the ER,
so the power mismatch estimates of ICUs neighboring the ER at each iteration are set to $0$ in (\ref{itpm1}).
Here, in (\ref{itpm1})-(\ref{pmguji}), the function of the ER as an information intermediary is explicitly shown and that
as an energy intermediary is implicitly embodied.

\vskip 0.2cm

For the above distributed algorithm, we have the following assumptions.

\begin{assumption}
\label{A1}{\rm The digraph $\overline{\mathcal{G}}$ contains a spanning tree with node $0$ as its root node.}
\end{assumption}

\begin{assumption}
\label{A1prime}
{\rm For any given node $i\in \{1,2,...,N\}$ of the undirected subgraph $\mathcal{G}$ of $\overline{\mathcal{G}}$, there is a
path from node $i$ to the root node $0$.}
\end{assumption}

\begin{assumption}
\label{A2}{\rm The algorithm step $\epsilon_i\in(0,\frac{1}{\sum_{j=0}^{N} a_{ij}})$.}
\end{assumption}

\begin{assumption}
\label{A3}{\rm The algorithm step $\mu\in(0,\frac{1}{\max_{i=1,2...,N}\sum_{j=1}^{N} a_{ij}})$.}
\end{assumption}

\vskip 0.2cm

A digraph satisfying Assumptions \ref{A1} and \ref{A1prime} is shown in Fig. 2.

\vskip 0.2cm
\begin{figure}[!t]
\centering
\includegraphics[width=2in]{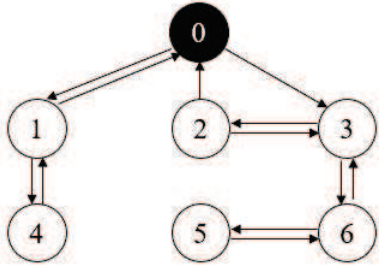}
\caption{A digraph satisfying Assumptions \ref{A1}-\ref{A1prime}.}
\label{figg1}
\end{figure}

\vskip 0.2cm

\emph{Remark 4}:
Assumption \ref{A1} ensures that the electricity price information of the distribution system can be transmitted from the ER to each ICU. Assumption \ref{A1prime} ensures that all ICUs can transmit the estimated average power mismatch of all bus nodes to the ER, and then the ER can calculate the incremental active power  exchanged between the distribution system and the microgrid.

\vskip 0.2cm

Denote $ {Y}(k)$ $=$ $[y_1(k),$ $y_2(k)$, $...$, $y_N(k)]^T$, $\Delta {P}(k)$ $=$ $[\Delta P_1(k)$, $\Delta P_2(k)$, $...$, $\Delta P_N(k)]^T$, $ {P}(k)$ $=$ $[P_1(k)$, $P_2(k)$, $...$, $P_N(k)]^T$, $\Delta \widehat{ {P}}(k)$ $=$ $[ \Delta\widehat{P}_1(k)$, $\Delta\widehat{P}_2(k)$, $...$, $\Delta\widehat{P}_N(k)]^{T}$, $ {\lambda}(k)$ $=$ $[\lambda_1(k),$ $\lambda_2(k)$, $...$, $\lambda_N(k)]^T$, $ {\epsilon}=diag\{\epsilon_1,$ $\epsilon_2,$ $...,$ $\epsilon_N\}$. For the convergence of the distributed ED algorithm (\ref{itlambda1})-(\ref{pmguji}), we have the following theorem.

\vskip 0.2cm
\begin{theorem}   
\label{oadd1}
For the algorithm (\ref{itlambda1})-(\ref{pmguji}), if Assumptions \ref{A1}-\ref{A3} hold, then
\begin{equation*}
\lim_{k\to\infty}\lambda_i(k)=\lambda_0, \lim_{k\to\infty}P_i(k)=P_i^*, \lim_{k\to\infty}P_{MG}(k)=P_{MG}^*,
\end{equation*}
where $P_i^*$ is given by (\ref{jizhongadd1}), and $P_{MG}^*$ is given by (\ref{jizhongm}). This means that the incremental cost with penalty factor of each DG converges to the electricity price of the distribution system asymptotically,  the active power generation of each DG is asymptotically optimal, the microgrid system achieves the balance of power supply and demand, and thus, the optimal ED is achieved asymptotically.
\end{theorem}
\vskip 0.2cm

\emph{Proof}:
Without loss of generality, we assume that the nodes $\{1,2,...,M\}$ can send information to the ER directly which means that $a_{0i}=1$, $i\in\{1,2,...,M\}$.

Rewrite (\ref{itlambda1}) in a compact form, then we get
\begin{equation*}
{\lambda}(k+1)-\lambda_0=[ {I_N}- \epsilon( {L_{\mathcal{G}}}+ {\mathcal{A}_{*0}})] ({\lambda}(k)-\lambda_0)
\end{equation*}
The equation (\ref{itlambda1}) is a standard leader-following consensus algorithm (\cite{Liu2008Controllability}), which is a special case of distributed consensus
algorithms with digraphs (\cite{RenBeardBook}, \cite{Olfati-SaberFaxMurray2007}). If Assumptions \ref{A1} and \ref{A2} hold, then by Theorem 2.20 in \cite{RenBeardBook} or Theorem 2 in \cite{Olfati-SaberFaxMurray2007},
all the eigenvalues of ${I_N}- \epsilon( {L_{\mathcal{G}}}+ {\mathcal{A}_{*0}})$ are inside the unit disk of the complex plane. Then we get
\begin{equation*}
\lim_{k\to\infty}\lambda_i(k)=\lambda_0,\ i=1,2,...,N.
\end{equation*}
Then by the above and (\ref{itpi1}), we have
\begin{equation}
\label{pkpk}
\lim_{k\to\infty}P_i(k)=P_i^*,\ i=1,2,...,N,
\end{equation}
Rewrite (\ref{itpm1}) in a compact form, then we get
\begin{eqnarray}
\label{aladdadd11}       
\left\{                  
\begin{array}{lll}     
     {Y}(k+1)=( {I_N}-\mu L_{\mathcal{G}})\Delta\widehat{ {P}}(k)+ {\xi}(k),
    \\
   \Delta \widehat{ {P}}(k+1)=( {I_N}- {C}) {Y}(k+1),
\end{array}
\right.k=0,1,2...,               
\end{eqnarray}
where $ {\xi}(k)=\Delta  {P}(k+1)-\Delta  {P}(k)$ and $ {C}= {\mathcal{A}_{0*}}$.

From (\ref{pkpk}), we know that $\lim_{k\to\infty} {\xi}(k)= {0}_{N\times 1}$.
Since $\mathcal{G}$ is undirected, $ {1}_N^T L_{\mathcal{G}}=0$. Then by (\ref{pmguji}) and (\ref{aladdadd11}), we have
\begin{equation*}
\begin{split}
  &\quad {1}_N^T[\Delta \widehat{ {P}}(k+1)-\Delta  {P}(k+1)]+P_{MG}(k+1)\\&= {1}_N^T[( {I_N}-\mu   {L}_{\mathcal{G}})\Delta \widehat{ {P}}(k)-\Delta  {P}(k)]+P_{MG}(k)\\
  &=...= {1}_N^T[\Delta \widehat{ {P}}(0)-\Delta  {P}(0)]+P_{MG}(0),\ k=0,1,2...,
\end{split}
\end{equation*}
which together with  ${1}_N^T[\Delta \widehat{ {P}}(0)-\Delta  {P}(0)]+P_{MG}(0)=0$ leads to
\begin{equation}
\label{aladdyyy}
  \sum_{i=1}^N\Delta \widehat{P}_i(k)+P_{MG}(k)=\sum_{i=1}^N\Delta P_i(k),\ k=0,1,2...
\end{equation}
From (\ref{aladdadd11}), we have
\begin{equation*}
  \Delta \widehat{ {P}}(k+1)=( {I}_N- {C})\Big[( {I_N}-\mu   {L}_{\mathcal{G}})\Delta \widehat{ {P}}(k)+ {\xi}(k)\Big], k=0,1,2...
\end{equation*}
From Assumption \ref{A1prime}, we know that for any $i=M+1, M+2,...,N$, there is $j\in\{1,2,...,M\}$, such that there is a path from $j$ to $i$. Then by Assumption \ref{A3} and Lemma \ref{oaddhxxx}, it is known that the eigenvalues of $( {I}_N- {C})( {I_N}-\mu   {L}_{\mathcal{G}})$ are all inside the unit disk of the complex plane. Then by $\lim_{k\to\infty} {\xi}(k)= {0}_{N\times 1}$, we have
\begin{equation*}
 \lim_{k\to\infty} \Delta \widehat{P}_i(k)=0, i=1,2,...,N.
 \end{equation*}
This together with (\ref{aladdyyy}), (\ref{pkpk}) and (\ref{jizhongm}) gives
\begin{equation*}
  \lim_{k\to\infty}P_{MG}(k)=\lim_{k\to\infty}\sum_{i=1}^N\Delta P_i(k)=P_{MG}^*,
\end{equation*}
that is, the microgrid system achieves the balance of power supply and demand asymptotically.
\hfill $\blacksquare$
\vskip 0.2cm

\emph{Remark 5}:
 If Assumptions \ref{A1} and \ref{A2} holds, then all the eigenvalues of ${I_N}- {\epsilon}( {L_{\mathcal{G}}}+ {\mathcal{A}_{*0}})$ are inside the unit disk of the complex plane (\cite{Olfati-SaberFaxMurray2007}-\cite{Seneta2006Nonnegative}), which ensures the convergence of the algorithm (\ref{itlambda1}).
According to Lemma \ref{oaddhxxx}, if Assumption \ref{A3} holds, all the eigenvalues of $(I_N-C)(I_N-\mu L_{\mathcal{G}})$ are inside the unit disk of the complex plane,
which ensures the convergence of the algorithm (\ref{itpm1}).

\vskip 0.2cm

\emph{Remark 6}: In \cite{wood1996power},
a quadratic transmission loss model is given by
\begin{equation}
\label{sunhaomox}
  P_L(P_1,P_2,...,P_N)=\sum_{i=1}^N\sum_{j=1}^NB_{ij}P_iP_j,
\end{equation}
where $B=[B_{ij}]_{N\times N}$ is a positive semi-definite matrix. This transmission loss model is simplified from the more general model known as Kron's loss
formula
\begin{equation*}
  P_L(P_1,P_2,...,P_N)=\sum_{i=1}^N\sum_{j=1}^NB_{ij}P_iP_j+\sum_{i=1}^NB_{0i}P_i+B_{00},
\end{equation*}
whose linear and constant terms are neglected in (\ref{sunhaomox}).
Noting that the diagonal elements are generally much larger than the non-diagonal elements in the loss matrix of $B$ (\cite{wood1996power}), the more simplified transmission loss model $P_L(P_1,...,P_N)=\sum_{i=1}^{N}B_iP_{i}^2$ is also widely used in the literature (\cite{Li2019Consensus}, \cite{Soliman2011Modern}, \cite{Yalcinoz1998Neural}).

For the problem (\ref{main}), if the quadratic transmission loss model (\ref{sunhaomox}) is used, then, similarly, from the KKT necessity condition it is known that
\begin{eqnarray*}       
\left\{                  
\begin{array}{lll}     
    \lambda^*=\frac{(P_{i}^*-\alpha_i)}{\beta_i(1-2B_{ii}P_i^*-2\sum_{j=1,j\neq i}^NB_{ij}P_j)},
     \\
    \lambda^*=\lambda_0,
\end{array}           
\right.i\notin\Omega(P_i^*)\cup\Gamma(P_i^*),            
\end{eqnarray*}
Denote
$P^{*}$$=$$[P_1^*,$$P_2^*,$$...,$$P_N^*]^T$, $Z$$=$$\left[
\begin{matrix}
 \frac{\alpha_1}{\beta_1}+\lambda_0 \      \frac{\alpha_2}{\beta_2}+\lambda_0  \     \cdots \ \frac{\alpha_N}{\beta_N}+\lambda_0
\end{matrix}
\right]^T$ and
\begin{equation*}
 {X=\left[
\begin{matrix}
 \frac{1}{\beta_1}+2\lambda_0B_{11}      & 2\lambda_0B_{12}      & \cdots & 2\lambda_0B_{1N}      \\
2\lambda_0B_{21}      & \frac{1}{\beta_2}+2\lambda_0B_{22}      & \cdots & 2\lambda_0B_{2N}      \\
 \vdots & \vdots & \ddots & \vdots \\
2\lambda_0B_{N1}      & 2\lambda_0B_{N2}      & \cdots & \frac{1}{\beta_N}+2\lambda_0B_{NN}      \\
\end{matrix}
\right]}.
\end{equation*}
If $i\notin\Omega(P_i^*)\cup\Gamma(P_i^*)$, $\forall$ $i=1,2,..., N$, then $XP^*=Z$. Noting that $B$ is a semi-positive matrix,  $\lambda_0>0$, $\beta_i>0$, $i=1,2,...,N$, we know that $X$ is a positive definite matrix, and
$P^*=X^{-1}Z$. Unlike (\ref{jizhongadd1}), even if all $P_{i}^{*}$ are not at the border, the optimal active power of each DG also depends on the cost parameters of all others and all the parameters of the loss matrix $B$. For this case, designing a distributed algorithm to compute the optimal solution is totally different from (\ref{itlambda1})-(\ref{pmguji}) and would merit more investigation in future.

\vskip 0.2cm



If the power system is dominated by the main grid and the microgrid mainly operates in the grid-connected mode, then the proposed algorithm (\ref{itlambda1})-(\ref{pmguji}) is effective. As more and more DGs and microgrids are added to the power system, an autonomous distributed ED algorithm covering both grid-connected and isolated modes of the microgrid should be considered. In the next section, based on (\ref{itlambda1})-(\ref{pmguji}), we will propose a new distributed ED algorithm. Although the algorithm requires a slightly stronger communication topology condition than the grid-connected algorithm (\ref{itlambda1})-(\ref{pmguji}), it covers both grid-connected and isolated modes of the microgrid, and can perform a smooth transition between both modes.

\section{Distributed Economic Dispatch Integrating Isolated and Grid-connected Modes}

\begin{figure}[!t]
\centering
\includegraphics[width=3in]{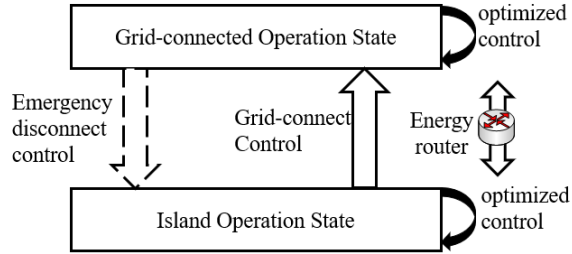}
\caption{The operation modes of microgrids.}
\label{fig2}
\end{figure}

As is well-known, the microgrid usually has two  operation modes, namely, isolated operation mode (island operation) and networked operation mode (grid-connected operation).  A microgrid should be able to perform a smooth transition between both modes to cope with emergencies in the main grid. For example, when a disaster occurs in the main grid, the microgrid switches to the isolated mode to avoid large-scale power outage, and the grid-connected mode is restored after the main grid becomes stable again.
For an Energy Internet, the operation modes of the microgrid are determined by the associated ER (\cite{cao2014energy}).
The two operation modes of a microgrid and their mutual transition are shown in Fig. \ref{fig2}.

In this section, we will design an autonomous distributed ED algorithm which integrates the two operation modes of the microgrid together. The characteristics of the algorithm lie in that the numerous ICUs (agents) which are not neighbors of the ER  do not need to know the operation mode of the microgrid, such that the all the agents of the microgrid can switch between the two operation modes autonomously.

The algorithm is divided into four parts. In the first part, each agent iterates based on the local information and obtains the incremental cost with penalty factor of its associated DG at time $k+1$.
\begin{equation}
\label{itlambda}
\begin{split}
\lambda_{i}(k+1)=\lambda_{i}(k)+\epsilon_i^\prime\Big[\sum_{j=1}^Na_{ij}(\lambda_j(k)-\lambda_i(k))+ga_{i0}(\lambda_0-\lambda_i(k))\Big]+\sigma(k)\Delta \widehat{P}_i(k),
\end{split}
\end{equation}
where $\epsilon_i^\prime>0$ is the step size of the  algorithm, $\lambda_i(0)$ is any given initial value, and $\sigma(k)>0$ is the feedback gain. If $0\in N_i$ and $g=1$, then $ga_{i0}>0$, which means that the ER transmits the electricity price information of the distribution system to its neighboring agents only when the microgrid is grid-connected. If $0\notin N_i$ or $g=0$, then $ga_{i0}=0$, which means that when the microgrid is in an isolated mode or although the whole microgrid is grid-connected, the non-neighboring agents of the ER do not need to know the electricity price of the distribution system.

\vskip 0.2cm

In the second part, each agent calculates the active power generated by each DG at time $k$ with
$\lambda_{i}(k)$.
\begin{equation}
\begin{split}
\label{itpi}    
P_{i}(k)=\Phi_i(\lambda_{i}(k))=\left\{                  
\begin{array}{lll}     
         \frac{\beta_i\lambda_i(k)+\alpha_i}{1+2B_i\beta_i\lambda_i(k)}, &\underline{P}_i\le \frac{\beta_i\lambda_i(k)+\alpha_i}{1+2B_i\beta_i\lambda_i(k)}\le\overline{P}_i, \\
     \overline{P}_{i},  &\frac{\beta_i\lambda_i(k)+\alpha_i}{1+2B_i\beta_i\lambda_i(k)}>\overline{P}_i,\\
     \underline{P}_{i},  &\frac{\beta_i\lambda_i(k)+\alpha_i}{1+2B_i\beta_i\lambda_i(k)}<\underline{P}_{i},
\end{array}           
\right.            
 \end{split}
\end{equation}
where $P_{i}(k)$ represents the active power generated by the $i$th DG at time $k$.
Similarly, for (\ref{itpi}), when $1+2B_i\beta_i\lambda_i(k)=0$, it is stipulated that  if $\beta_i\lambda_i(k)+\alpha_i>0$, then $P_{i}(k)=\overline{P}_{i}$, while  if $\beta_i\lambda_i(k)+\alpha_i<0$,  then $P_{i}(k)=\underline{P}_i$.

The third part of the algorithm consists of four iterations. Each agent estimates the average power mismatch of all bus nodes of the microgrid system through average consensus algorithm. During each iteration, each agent transmits its estimate to the ER, and then the ER calculates the incremental active power that each bus node needs to exchange with the distribution system.
\bna
\label{itpm}
y_i(k+1)&=&\Delta \widehat{P}_i(k)+\mu^\prime\Big[\sum_{j\in {N_i}}a_{ij}(\Delta \widehat{P}_j(k)-\Delta \widehat{P}_i(k))\Big]+\Delta P_i(k+1)-\Delta P_i(k),\\
\label{itpm2}
\Delta P_{Mi}(k+1)&=&a_{0i}gy_i(k+1),\\
\label{itpm23}
P_{Mi}(k+1)&=&g[P_{Mi}(k)+a_{i0}\Delta P_{Mi}(k+1)], \\
\label{itpm235}
\Delta \widehat{P}_i(k+1)&=&y_i(k+1)+a_{i0}[P_{Mi}(k)-P_{Mi}(k+1)],
\ena
where $\Delta P_i(k)$
is the power mismatch of the $i$th bus node, $\mu^\prime>0$ is the algorithm step size, and $\Delta \widehat{P}_i(k)$ is the local estimate of the average power mismatch of all buses with  $\sum_{i=1}^{N}\Delta \widehat{P}_i(0)=\sum_{i=1}^{N}\Delta P_i(0)$ and $P_{Mi}(0)=0$. Here, $\Delta P_{Mi}(k)$ represents the incremental active power that the $i$th bus node  needs to exchange with the distribution system at time $k$.

In the fourth part of the algorithm, the ER calculates the active power exchanged with the distribution system for the whole microgrid.
\begin{equation}
\label{pm}
P_{MG}(k)=\sum_{i=1}^NP_{Mi}(k),        
\end{equation}
where $P_{MG}(k)$ represents the active power exchanged with the distribution system.

\vskip 0.2cm

\emph{Remark 7}:
 The equations (\ref{itpm2}), (\ref{itpm23}) and (\ref{pm}) are performed by the ER. The equations (\ref{itpm2}) and (\ref{itpm23}) indicate that in the grid-connected mode, the power exchanged between the microgrid and the distribution system is  continuously accumulated
by the ER during each iteration. If $i\notin N_0$, then $P_{Mi}(k)\equiv0$, or for the isolated mode with $g=0$, $P_{Mi}(k)\equiv0$, $i=1,2,...,N$.

\vskip 0.2cm

\emph{Remark 8}:
The equation (\ref{itpm235}) together with (\ref{itpm2})-(\ref{itpm23}) means  that for the bi-directionally neighboring agents of the ER, there is direct power replenishment by the distribution system through the ER after each iteration, and so their estimates for average power mismatch are zeros. Noticing that the microgrid should perform a smooth transition between the grid-connected  and the isolated modes, (\ref{itpm235}) can match the power exchanged with the distribution system to the power mismatch of the microgrid system when the microgrid switches from the grid-connected mode to the isolated mode.

\vskip 0.2cm
For the proposed algorithm (\ref{itlambda})-(\ref{pm}), we have the following assumptions.

\begin{assumption}
\label{A4}{\rm The undirected subgraph $\mathcal{G}$ is connected and there is a node $i\in\{1,2,...,N\}$, such that $a_{i0}a_{0i}>0$.}
\end{assumption}

\begin{assumption}
\label{A5}{\rm The algorithm step $\epsilon_i^\prime\in(0,\frac{1}{\sum_{j=0}^{N} a_{ij}})$.}
\end{assumption}

\begin{assumption}
\label{A6}{\rm The algorithm step  $\mu^\prime$$\in$$(0,$$\frac{1}{\max_{i=1,2...,N}\sum_{j=1}^{N} a_{ij}})$.}
\end{assumption}

\begin{assumption}
\label{A7}{\rm The feedback gain $\sigma(k)>0$, $\lim_{k\rightarrow\infty}\sigma(k)=0$ and  $\sum_{k=0}^\infty\sigma(k)=\infty$.}
\end{assumption}

\vskip 0.2cm

A digraph satisfying Assumption \ref{A4} is shown in Fig. 4.

\vskip 0.2cm

\emph{Remark 9}:
Different from the algorithm (\ref{itlambda1}) which is only for the grid-connected mode, we add a feedback term $\sigma(k)\Delta \widehat{P}_i(k)$  in the algorithm (\ref{itlambda}). Without this term, if the microgrid is in the isolated mode, that is, $g=0$, then (\ref{itlambda}) becomes the average consensus algorithm, and all $\lambda_{i}(k)$, $i=1,2,...,N$  will converge to $\frac{1}{N}\sum_{i=1}^N\lambda_{i}(0)$
instead of $\lambda^{*\prime}$ in (\ref{jizhongnlam}).  On one hand, the algorithm (\ref{itlambda}) uses the local estimate $\Delta \widehat{P}_i(k)$ of the average power mismatch of all buses to drive $\lambda_{i}(k)$ away from $\frac{1}{N}\sum_{i=1}^N\lambda_{i}(0)$ when the microgrid is in the isolated mode, on the other hand, the vanishing feedback gain $\sigma(k)$ does not excessively block the function of the consensus term $\sum_{j=1}^Na_{ij}(\lambda_j(k)-\lambda_i(k))+ga_{i0}(\lambda_0-\lambda_i(k))$.

\vskip 0.2cm

\emph{Remark 10}:
If Assumption \ref{A4} holds, then Assumptions \ref{A1} and \ref{A1prime} hold. Here, Assumption \ref{A4} on the network graph is stronger than  Assumptions \ref{A1} and \ref{A1prime} for the algorithm (\ref{itlambda1})-(\ref{pmguji}) in the grid-connected operation. For a distributed ED algorithm covering both grid-connected and isolated modes of the microgrid, it is necessary that at least one ICU can transmit its own estimated average power mismatch to the ER as in (\ref{itpm2}), and receive the information that how much power is needed to be exchanged between its associated bus node and the distribution system calculated by the ER  as in (\ref{itpm23}). For this ICU, if the microgrid is switched to isolated mode from grid-connected mode at time $k+1$£¬then $P_{Mi}(k+1)=0$£¬and this ICU
get the information that how much power has been exchanged between its associated bus node and the distribution system  $P_{Mi}(k)$ as in (\ref{itpm235}), such that the estimate for the total power mismatch of the microgrid system is always equal to the real total power mismatch of the microgrid system no matter the microgrid is in grid-connected or isolated mode and no matter when mode switching happens. More details will be discussed in Theorem \ref{oadd1}.
\vskip 0.2cm


\begin{figure}[!t]
\centering
\includegraphics[width=2in]{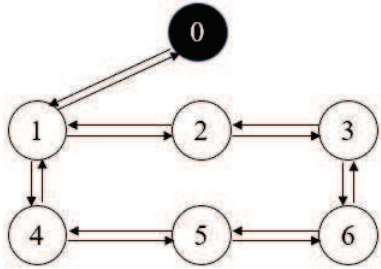}
\caption{A digraph satisfying Assumptions \ref{A1} and \ref{A4}.}
\label{figg2}
\end{figure}

Denote $ {Y}(k)$ $=$ $[y_1(k),$ $y_2(k),$ $...,$ $y_N(k)]^T,$ $\Delta {P}(k)$ $=$ $[\Delta P_1(k),$ $\Delta P_2(k),$ $...,$ $\Delta P_N(k)]^T,$ $ {P}(k)$ $=$ $[P_1(k),$ $P_2(k),$ $...,$ $P_N(k)]^T,$ $\Delta\widehat{ {P}}(k)$ $=$ $[\Delta\widehat{P}_1(k),$ $\Delta\widehat{P}_2(k),$..., $\Delta\widehat{P}_N(k)]^{T},$ $\Delta {P}_M(k)$ $=$ $[\Delta P_{M1}(k),$ $\Delta P_{M2}(k),$ $...,$ $\Delta P_{MN}(k)]^T,$ $ {P}_M(k)$ $=$ $[ P_{M1}(k),$ $P_{M2}(k),$ $...,$ $ P_{MN}(k)]^T;$ $\Phi( {\lambda}(k))=[\Phi_1(\lambda_1(k)),$ $\Phi_2(\lambda_2(k)),$ $...,$ $\Phi_N(\lambda_N(k))]^T,$ $ {B}=diag\{B_1,$ $B_2,$ $...,$ $B_N\},$ $ {\epsilon}^\prime=diag\{\epsilon_1^\prime,$ $\epsilon_2^\prime,$ $...,$ $\epsilon_N^\prime\}$. If the microgrid is grid-connected, then we have the following theorem.

\vskip 0.2cm

\begin{theorem}   
\label{oaddhhh}
For the algorithm (\ref{itlambda})-(\ref{pm}), if Assumptions \ref{A4}-\ref{A7} hold and $g=1$, then
\begin{equation*}
\lim_{k\to\infty}\lambda_i(k)=\lambda_0, \lim_{k\to\infty}P_i(k)=P_i^*, \lim_{k\to\infty}P_{MG}(k)=P_{MG}^*,
\end{equation*}
where $P_i^*$ is given by (\ref{jizhongadd1}), and $P_{MG}^*$ is given by (\ref{jizhongm}). This means that the incremental cost with penalty factor of each DG converges to the electricity price of the distribution system asymptotically,  the active power generation of each DG is asymptotically optimal, the microgrid system achieves the balance of power supply and demand, and thus, the optimal ED is achieved asymptotically.
\end{theorem}
\vskip 0.2cm

\emph{Proof}:
When $g=1$, the microgrid is in the grid-connected operation mode.

Without loss of generality, assume that the nodes $1,2,...,M^\prime$ are neighbors of the ER in bi-direction, which means that $a_{0i}=a_{i0}=1$, $i\in\{1,2,...,M^\prime\}$, and $ {\mathcal{A}_{*0}\mathcal{A}_{0*}}= {\mathcal{A}_{*0}\mathcal{A}_{*0}\mathcal{A}_{0*}}$.

Rewrite (\ref{itlambda})-(\ref{itpm235}) in a compact form, then we get
\begin{eqnarray}
\label{aladsss}       %
\left\{                  
\begin{array}{lll}     
      \begin{aligned}
   {\lambda}(k+1)&= {\epsilon}^\prime {\mathcal{A}_{*0}1}\lambda_0+[ {I_N}- {\epsilon}^\prime( {L_{\mathcal{G}}}+ {\mathcal{A}_{*0}})] {\lambda}(k)+\sigma(k)\Delta \widehat{ {P}}(k),
  \end{aligned}
    \\
     {P}(k)=\Phi( {\lambda}(k)),
    \\
    \begin{aligned}
    {Y}(k+1)&=( {I_N}-\mu^\prime {L_{\mathcal{G}}})\Delta \widehat{ {P}}(k)+\Delta {P}(k+1)-\Delta {P}(k),
   \end{aligned}
     \\
     \begin{aligned}
   \Delta  {P_M}(k+1)&= {\mathcal{A}_{0*}} {Y}(k),
   \end{aligned}
   \\
   \begin{aligned}
    {P_M}(k+1)&= {P_M}(k)+ {\mathcal{A}_{*0}}\Delta  {P_M}(k+1),
   \end{aligned}
   \\
   \begin{aligned}
   \Delta \widehat{ {P}}(k+1)&= {Y}(k+1)+ {\mathcal{A}_{*0}}[ {P_M}(k)- {P_M}(k+1)].
   \end{aligned}
\end{array}
\right.             
\end{eqnarray}
Denote $ {C^\prime}= {\mathcal{A}_{*0}\mathcal{A}_{0*}}$. From (\ref{aladsss}) and
$ {C^\prime}= {\mathcal{A}_{*0}\mathcal{A}_{*0}\mathcal{A}_{0*}}$, we have
\begin{eqnarray}
\label{aladsssddd}       %
\left\{                  
\begin{array}{lll}     
    \begin{aligned}
   \Delta \widehat{ {P}}(k+1)&=( {I_N}- {C^\prime}) {Y}(k),
   \end{aligned}
     \\
     \begin{aligned}
    {P_M}(k+1)- {P_M}(k)&= {C^\prime} {Y}(k),
   \end{aligned}
\end{array}
\right.             
\end{eqnarray}
From (\ref{aladsssddd}) and (\ref{aladsss}), we have
\begin{equation}
\label{xcvb}
\begin{split}
\Delta \widehat{ {P}}(k+1)&=( {I_N}- {C^\prime})( {I_N}-\mu^\prime {L_{\mathcal{G}}})\Delta \widehat{ {P}}(k)+( {I_N}- {C^\prime})(\Delta {P}(k+1)-\Delta {P}(k)).
\end{split}
\end{equation}
By the definition of $\Delta {P}(k)$, we have
\begin{equation}
\begin{split}
\label{pmbPkk}
\underset{k\geq0}{\sup}  \|\Delta {P}(k+1)-\Delta {P}(k)\|\leq2\underset{k\geq0}{\sup}\|\Delta {P}(k)\|\leq2\sqrt{N}\underset{i=1,2,...,N}{\max}(P_{Di}-\underline{P}_{i}-B_i\underline{P}^2_{i}).\\
\end{split}
\end{equation}
From Assumption \ref{A4}, we know  that for any $i=M^\prime+1, M^\prime+2,...,N$, there is $j\in\{1,2,...,M^\prime\}$, so that there is a path from $j$ to $i$. Then by Assumption \ref{A6} and Lemma \ref{oaddhxxx}, we get that the eigenvalues of $( {I}_N- {C}^\prime)( {I_N}-\mu^\prime   {L}_{\mathcal{G}})$ are all inside the unit disk. This together with (\ref{xcvb}) and (\ref{pmbPkk}) gives
\begin{equation}
\label{deltap}
\underset{k\geq0}{\sup} \| {\Delta \widehat{P}}(k)\|<\infty.
\end{equation}
From Assumptions \ref{A4} and \ref{A5}, we know  that the eigenvalues of $ {I_N}- {\epsilon}^\prime( {L_{\mathcal{G}}}+ {\mathcal{A}_{*0}})$ are all inside the unit disk of the complex plane. By (\ref{aladsss}), we get
$ {\lambda}(k+1)- {1}\lambda_0
=[ {I_N}- {\epsilon}^\prime( {L_{\mathcal{G}}}+ {\mathcal{A}_{*0}})]( {\lambda}(k)- {1}\lambda_0)+\sigma(k)\Delta \widehat{ {P}}(k).$

Then by Assumption \ref{A7} and (\ref{deltap}), we have $\lim_{k\rightarrow\infty}(\lambda_i(k)-\lambda_0)=0,\ i=1,2,...,N$.
This together with (\ref{itpi}) leads to
\begin{equation}
\label{pkpk22}
\lim_{k\to\infty}P_i(k)=P_i^*,,\ i=1,2,...,N,
\end{equation}
where $P_i^*$ is given by (\ref{jizhongadd1}). Then from the above and the definition of $\Delta {P}(k)$, we have
\begin{equation*}
\lim_{k\rightarrow\infty}(\Delta {P}(k+1)-\Delta {P}(k))= {0}.
\end{equation*}
This together (\ref{xcvb}) and Lemma \ref{oaddhxxx} leads to
\begin{equation}
\label{deltappp}
 \lim_{k\rightarrow\infty}\Delta \widehat{ {P}}(k)= {0}.
\end{equation}
From (\ref{aladsssddd}), (\ref{aladsss}) and (\ref{pm}), noticing that $ {1}_N^{T}L_{\mathcal{G}}=0$, we get
\bna
\label{deltapppbingwang}
&& {1}_N^T[\Delta \widehat{ {P}}(k+1)-\Delta  {P}(k+1)]+P_{MG}(k+1)\cr
&=& {1}_N^T[\Delta \widehat{ {P}}(k+1)+ {P_M}(k+1)- {P_M}(k)+ {P_M}(k)-\Delta  {P}(k+1)]\cr
&=& {1}_N^T[( {I_N}-\mu^\prime {L_{\mathcal{G}}})\Delta \widehat{ {P}}(k)-\Delta  {P}(k)]+ {1}_N^T {P_M}(k)\cr
&=& {1}_N^T[\Delta \widehat{ {P}}(k)-\Delta  {P}(k)]+ {1}_N^T {P_M}(k)\cr
&&...\cr
&=& {1}_N^T[\Delta \widehat{ {P}}(0)-\Delta  {P}(0)]+ {1}_N^T {P_M}(0)\cr
&=& {1}_N^T[\Delta \widehat{ {P}}(0)-\Delta  {P}(0)]+P_{MG}(0),\ k=0,1,2...
\ena
Then from ${1}_N^T\Delta\widehat{ {P}}(0)={1}_N^T\Delta{P}(0)$, $P_{Mi}(0)=0$ and the definition of $\Delta{P}(k)$, we have
\begin{equation*}
  \sum_{i=1}^N\Delta \widehat{P}_i(k)+P_{MG}(k)+\sum_{i=1}^NP_i(k)=\sum_{i=1}^N P_{Di}+\sum_{i=1}^NB_iP^2_i(k),
\end{equation*}
which together with (\ref{deltappp}) and (\ref{pkpk22}) leads to
\begin{equation*}
 \lim_{k\to\infty}P_{MG}(k)=\sum_{i=1}^N P_{Di}+\sum_{i=1}^{N}B_i(P_{i}^{*})^2-\sum_{i=1}^N P_{i}^*=P_{MG}^*.
\end{equation*}

\hfill $\blacksquare$

\vskip 0.2cm

If the microgrid is in the isolated mode, we have the following theorem.

\vskip 0.2cm

\begin{theorem}   
\label{oaddfff}
For the algorithm (\ref{itlambda})-(\ref{itpm235}), if Assumptions \ref{A4}-\ref{A7} hold and $g=0$, then for any $i,j\in{1,2,...,N},i\neq j$, we have
\bna
\label{equallambdaaddadd1}
\lim_{k\to\infty}(\lambda_i(k)-\lambda_j(k))=0,\ i,j=1,2,...,N,\cr
\underset{k\geq0}{\sup}|\Delta \widehat{P}_i(k)|<\infty, i=1,2,...,N.
\ena
That is,  for all DGs, the incremental costs with penalty factor tend to be equal asymptotically and the estimates of all ICUs for the average power mismatch are bounded.
\end{theorem}
\vskip 0.2cm

\emph{Proof}:
If $g=0$, then from (\ref{itpm23}), it follows that $P_{Mi}(k)\equiv0$. Then by (\ref{pm}), we have $P_{MG}(k)\equiv0$.

Rewrite (\ref{itlambda})-(\ref{itpm235}) in a compact form, then we get
\bna
\label{aladhhhh}
 {\lambda}(k+1)&=&( {I_N}- {\epsilon}^\prime {L_{\mathcal{G}}}) {\lambda}(k)+\sigma(k)\Delta \widehat{ {P}}(k),\\
 {P}(k)&=&\Phi( {\lambda}(k)),\\
\label{aladhhh2}
\Delta \widehat{ {P}}(k+1)&=&( {I_N}-\mu^\prime {L_{\mathcal{G}}})\Delta \widehat{ {P}}(k)+\Delta {P}(k+1)-\Delta {P}(k),
\ena
Denote $\delta_{ {\lambda}}(k)=( {I_N}- {J}_N) {\lambda}(k),\delta_{\widehat{ {P}}}(k)=( {I_N}- {J}_N)\Delta \widehat{ {P}}(k)$. From (\ref{aladhhh2}), we have
\bna
\label{deltaptaddadd1}
&&\hspace*{-0.3cm}\delta_{\widehat{ {P}}}(k)\cr
\hspace*{-0.3cm}&=&\hspace*{-0.3cm}( {I_N}-\mu^\prime {L_{\mathcal{G}}})\delta_{\widehat{ {P}}}(k-1)+( {I_N}- {J}_N)(
\Delta {P}(k)-\Delta {P}(k-1))\cr
\hspace*{-0.3cm}&=&\hspace*{-0.3cm}( {I_N}-\mu^\prime {L_{\mathcal{G}}})^k\delta_{\widehat{ {P}}}(0)+\sum^{k-1}_{j=0}( {I_N}-\mu^\prime {L_{\mathcal{G}}})^{k-1-j}( {I_N}- {J}_N)(\Delta {P}(j+1)-\Delta {P}(j)).
\ena
Then by Assumptions \ref{A4} and \ref{A6} and Theorem 4.2 in \cite{Seneta2006Nonnegative}, we know that
\begin{equation}
\label{xuyao1}
\lim_{k\to\infty}( {I_N}-\mu^\prime {L_{\mathcal{G}}})^k= {J}_N,
\end{equation}
and
\bna
\label{expoconveraddadd1}
\|( {I_N}-\mu^\prime {L_{\mathcal{G}}})^{k}- {J}_N\|\leq c_1\rho_1^k,
\ena
where $\rho_1\in(0,1)$ and $c_1>0$ are both non-negative constants.
From (\ref{xuyao1}) and the definition of $\delta_{\widehat{ {P}}}(k)$, we get
\bna
\label{deltafirstpartconvergence}
\lim_{k\to\infty}( {I_N}-\mu^\prime {L_{\mathcal{G}}})^k\delta_{\widehat{ {P}}}(0)= {0}.
\ena
From the properties of the Laplacian matrix $L_{\mathcal{G}}$, it is known that
\bna
\label{xuyao2}
&&( {I_N}-\mu^\prime {L_{\mathcal{G}}})^{k-1-j} {J}_N\cr
&=&( {I_N}-\mu^\prime {L_{\mathcal{G}}})^{k-j-2}( {J}_N-\mu^\prime {L_{\mathcal{G}}} {J}_N)\cr
&=&( {I_N}-\mu^\prime {L_{\mathcal{G}}})^{k-j-2} {J}_N\cr
&=&...\cr
&=& {J}_N.
\ena
This together with (\ref{expoconveraddadd1}) leads to
\ban
&&\hspace*{-0.3cm}\|\sum^{k-1}_{j=0}( {I_N}-\mu^\prime {L_{\mathcal{G}}})^{k-1-j}( {I_N}- {J}_N)(\Delta {P}(j+1)- {P}(j))\|\cr
\hspace*{-0.3cm}&=&\hspace*{-0.3cm}\|\sum^{k-1}_{j=0}(( {I_N}-\mu^\prime {L_{\mathcal{G}}})^{k-1-j}- {J}_N)(\Delta {P}(j+1)-\Delta {P}(j))\|\cr
\hspace*{-0.3cm}&\leq&\hspace*{-0.3cm}\sum^{k-1}_{j=0}\|( {I_N}-\mu^\prime {L_{\mathcal{G}}})^{k-1-j}- {J}_N\|\|(\Delta {P}(j+1)-\Delta {P}(j))\|\cr
\hspace*{-0.3cm}&\leq&\hspace*{-0.3cm}\sum^{k-1}_{j=0}c_1\rho_1^{k-1-j}2\underset{j\geq0}{\sup}\|\Delta {P}(j)\|\cr
\hspace*{-0.3cm}&\leq&\hspace*{-0.3cm}2c_1\underset{j\geq0}{\sup}\|\Delta {P}(j)\|\sum^{\infty}_{j=0}\rho_1^j\cr
\hspace*{-0.3cm}&\leq&\hspace*{-0.3cm}\frac{2c_1\sqrt{N}\underset{i=1,2,...,N}{\max}(P_{Di}-\underline{P}_{i}-B_i\underline{P}^2_{i})}{1-\rho_1}.
\ean
Then from (\ref{deltaptaddadd1}) and (\ref{deltafirstpartconvergence}), we have
\begin{equation}
\label{deltapkaddadd1}
\underset{k\geq0}{\sup}\|\delta_{\widehat{ {P}}}(k)\|<\infty.
\end{equation}
From (\ref{aladhhh2}) and $ {1}_N^{T}L_{\mathcal{G}}=0$, it is known that
\begin{equation*}
\begin{split}
  &\quad {1}_N^T[\Delta \widehat{ {P}}(k+1)-\Delta {P}(k+1)]\\
  &= {1}_N^T[( {I_N}-\mu^\prime {L_{\mathcal{G}}})\Delta \widehat{ {P}}(k)-\Delta  {P}(k)]\\
  &= {1}_N^T[\Delta \widehat{ {P}}(k)-\Delta  {P}(k)]\\
  &...\\
  &= {1}_N^T[\Delta \widehat{ {P}}(0)-\Delta  {P}(0)]\\
  &=0,\ k=0,1,2...,
\end{split}
\end{equation*}
which means
\begin{equation}
\label{ppbaaddaddadd1}
 {1}_N^T\Delta \widehat{ {P}}(k)= {1}_N^T\Delta {P}(k),\ k=0,1,2...
\end{equation}
Then from the above and the definition of $\delta_{\widehat{ {P}}}(k)$, we get
\begin{equation}
\begin{split}
\label{deltapk}
\Delta \widehat{ {P}}(k)&= {J}_N\Delta \widehat{ {P}}(k)+\delta_{\widehat{ {P}}}(k)=\frac{1}{N} {1}_N{1}_N^T\Delta \widehat{ {P}}(k)+\delta_{\widehat{ {P}}}(k)={J}_N\Delta {P}(k)+\delta_{\widehat{ {P}}}(k), \ k=0,1,2...
\end{split}
\end{equation}
From (\ref{itpi}), we know that $\sup_{k\geq0}\|\Delta {P}(k)\|< {\infty}$. Then by (\ref{deltapkaddadd1}), we get
\begin{equation*}
\underset{k\geq0}{\sup}\|\Delta \widehat{ {P}}(k)\|< {\infty}.
\end{equation*}
That is, $\underset{k\geq0}{\sup}|\Delta \widehat{P}_i(k)|<\infty$, $i=1,2,...,N$. By (\ref{aladhhhh}) and the definition of $\delta_{ {\lambda}}(k)$, we have
\bna
\label{deltalambdaaddadd1}
&&\hspace*{-0.3cm}\delta_{ {\lambda}}(k)\cr
\hspace*{-0.3cm}&=&\hspace*{-0.3cm}( {I_N}- {\epsilon}^\prime {L_{\mathcal{G}}})( {I_N}- {J}_N) {\lambda}(k-1)+\sigma(k-1)( {I_N}- {J}_N)\Delta \widehat{ {P}}(k-1)\cr
\hspace*{-0.3cm}&=&\hspace*{-0.3cm}( {I_N}- {\epsilon}^\prime {L_{\mathcal{G}}})\delta_{ {\lambda}}(k-1)+\sigma(k-1)\delta_{\widehat{ {P}}}(k-1)\cr
\hspace*{-0.3cm}&=&\hspace*{-0.3cm}( {I_N}- {\epsilon}^\prime {L_{\mathcal{G}}})^k\delta_{ {\lambda}}(0)
+\sum^{k-1}_{j=0}( {I_N}- {\epsilon}^\prime {L_{\mathcal{G}}})^{k-1-j}\sigma(j)\delta_{\widehat{ {P}}}(j), k=0,1,2,...
\ena
From Assumptions \ref{A4} and \ref{A5}, similarly to (\ref{xuyao1}) and (\ref{expoconveraddadd1}), it is known that
\begin{equation}
\label{xuyao133add}
\lim_{k\to\infty}( {I_N}- {\epsilon}^\prime {L_{\mathcal{G}}})^k= {J}_N,
\end{equation}
and
$\|( {I_N}- {\epsilon}^\prime {L_{\mathcal{G}}})^{k}- {J}_N\|\leq c_2\rho_2^k$,
where $\rho_2\in(0, 1)$, $c_2>0$ are both non-negative constants.
From (\ref{xuyao133add}), we get
\bna
\label{deltapconveraddaddadd1}
\lim_{k\to\infty}( {I_N}- {\epsilon}^\prime {L_{\mathcal{G}}})^k\delta_{\widehat{ {P}}}(0)= {0}.
\ena
Similarly to (\ref{xuyao2}), we have
$( {I_N}- {\epsilon}^\prime {L_{\mathcal{G}}})^{k-1-j} {J}_N= {J}_N,\ j=0,1,...,k-1$.
This together with Assumption \ref{A7}, (\ref{xuyao133add}), (\ref{deltalambdaaddadd1}) and (\ref{deltapkaddadd1}) leads to
\begin{equation*}
\begin{split}
&\quad\|\delta_{ {\lambda}}(k)\|\\
&\leq\|( {I_N}- {\epsilon}^\prime {L_{\mathcal{G}}})^k\delta_{ {\lambda}}(0)\|
+\sum^{k-1}_{j=0}( {I_N}- {\epsilon}^\prime {L_{\mathcal{G}}})^{k-1-j}\sigma(j)\delta_{\widehat{ {P}}}(j)\|\\
&=o(1)+\|\sum^{k-1}_{j=0}(( {I_N}- {\epsilon}^\prime {L_{\mathcal{G}}})^{k-1-j}- {J}_N+ {J}_N)\sigma(j)\delta_{\widehat{ {P}}}(j)\|\\
&=o(1)+\sum^{k-1}_{j=0}\|( {I_N}- {\epsilon}^\prime {L_{\mathcal{G}}})^{k-1-j}- {J}_N\|\sigma(j)\|\delta_{\widehat{ {P}}}(j)\|\\
&\leq o(1)+\sum^{k-1}_{j=0}c_2\rho_2^{k-1-j}\sigma(j)\|\delta_{\widehat{ {P}}}(j)\|\\
&\leq o(1)+\sup_{j\geq0}\|\delta_{\widehat{ {P}}}(j)\|\sum^{k-1}_{j=0}c_2\rho_2^{k-1-j}\sigma(j)=o(1),\ k\to\infty,\\
\end{split}
\end{equation*}
which implies (\ref{equallambdaaddadd1}), that is,  for all DGs, the incremental costs with penalty factor tend to be equal asymptotically.
\hfill $\blacksquare$
\vskip 0.2cm
Theorem \ref{oaddfff} shows that in the isolated operation mode, the algorithm (\ref{itlambda})-(\ref{pm}) ensures that for all DGs, the incremental costs with penalty factor tend to be equal asymptotically. Numerical simulation shows that for all DGs, the incremental costs with penalty factor will converge to a common value (see Section V.B as shown in Figure \ref{fig6}.a.). It can be proved that for all DGs, the incremental costs with penalty factor converge to the same value, the microgrid system achieves optimal ED and the balance of power supply and demand asymptotically under the assumption that for all DGs, the incremental costs with penalty factor converge. It needs far more investigation to remove this assumption and remains as an interesting open problem. We have the following proposition.

\vskip 0.2cm

\newtheorem{proposition}{Proposition}
\begin{proposition}   
\label{oaddddd}
For the algorithm (\ref{itlambda})-(\ref{pm}), suppose that Assumptions \ref{A4}-\ref{A7} hold and $g=0$. If $\{\lambda_i(k),k=0,1,...\}$, $i=1,2,...,N$ converge, then
\bna
\label{optimalisolatedaddadd1}
\lim_{k\to\infty}\lambda_i(k)=\lambda^{*\prime}, \lim_{k\to\infty}P_i(k)=P_i^{*\prime}, \lim_{k\rightarrow\infty}\Delta \widehat{P}_i(k)=0,
\ena
where $\lambda^{*\prime}$ and $P_i^{*\prime}$ are given by (\ref{jizhongnlam}) and (\ref{jizhongn}), respectively. Namely, the microgrid system achieves optimal ED and the balance of power supply and demand  asymptotically.
\end{proposition}
\vskip 0.2cm

\emph{Proof}:
If Assumptions \ref{A4}-\ref{A7} hold and $g=0$, then by Theorem \ref{oaddfff}, we have $\lim_{k\to\infty}(\lambda_i(k)-\lambda_j(k))=0$, $i,j\in{1,2,...,N}$, $i\neq j$.

From (\ref{deltaptaddadd1}), (\ref{expoconveraddadd1}) and (\ref{deltafirstpartconvergence}), we have
\bna
\label{dealt11}
&&\|\delta_{\widehat{ {P}}}(k)\|\cr
&=&\|( {I_N}-\mu^\prime {L_{\mathcal{G}}})^k\delta_{\widehat{ {P}}}(0)+\sum^{k-1}_{j=0}(( {I_N}-\mu^\prime {L_{\mathcal{G}}})^{k-1-j}- {J}_N)(\Delta {P}(j+1)-\Delta {P}(j))\|\cr
&\leq&\|( {I_N}-\mu^\prime {L_{\mathcal{G}}})^k\delta_{\widehat{ {P}}}(0)\|+\sum^{k-1}_{j=0}\|( {I_N}-\mu^\prime {L_{\mathcal{G}}})^{k-1-j}- {J}_N\|\|(\Delta {P}(j+1)-\Delta {P}(j))\|\cr
&\leq&o(1)+\sum^{k-1}_{j=0}c_1\rho_1^{k-1-j}\|(\Delta {P}(j+1)-\Delta {P}(j))\|.
\ena
If $\{\lambda_i(k), k=0,1,...\}$, $i=1,2,...,N$ converge, then from (\ref{itpi}), it follows that $\{ {P}(k)$, $k=0,1,...\}$ converges. Then by the definition of $\Delta P_i(k)$, we have $\{\Delta P_i(k)$, $k=0,1,...\}$, $i=1,2,...,N$ converge. Thus, for any given $\varepsilon>0$, there is a positive integer $L$, such that $\|\Delta  {P}(k+1)-\Delta {P}(k)\|\leq\varepsilon$, $k\geq L$. This implies that
\ban
&&\hspace*{-0.3cm}\sum^{k-1}_{j=0}c_1\rho_1^{k-1-j}\|(\Delta {P}(j+1)-\Delta {P}(j))\|\cr
\hspace*{-0.3cm}&=&\hspace*{-0.3cm}\sum^{L-1}_{j=0}c_1\rho_1^{k-1-j}\|(\Delta {P}(j+1)-\Delta {P}(j))\|+\sum^{k-1}_{j=L}c_1\rho_1^{k-1-j}\|(\Delta {P}(j+1)-\Delta {P}(j))\|\cr
\hspace*{-0.3cm}&\leq&\hspace*{-0.3cm}\rho_1^{k-1}\sum^{L-1}_{j=0}c_1\rho_1^{-j}\|(\Delta {P}(j+1)-\Delta {P}(j))\|+\sum^{k-1}_{j=L}c_1\rho_1^{k-1-j}\varepsilon\cr
\hspace*{-0.3cm}&=&\hspace*{-0.3cm}\rho_1^{k-1}\sum^{L-1}_{j=0}c_1\rho_1^{-j}\|(\Delta {P}(j+1)-\Delta {P}(j))\|+\sum^{k-1-L}_{j=0}c_1\rho_1^{j}\varepsilon\cr
\hspace*{-0.3cm}&\leq&\hspace*{-0.3cm}\rho_1^{k-1}\sum^{L-1}_{j=0}c_1\rho_1^{-j}\|(\Delta {P}(j+1)-\Delta {P}(j))\|+\sum^{\infty}_{j=0}c_1\rho_1^{j}\varepsilon\cr
\hspace*{-0.3cm}&=&\hspace*{-0.3cm}o(1)+\frac{\varepsilon c_1}{1-\rho_1},\ k\to\infty,
\ean
which together with (\ref{dealt11}) gives
\bna
\label{deltapconveraddadd1}
\lim_{k\to\infty}\|\delta_{\widehat{ {P}}}(k)\|=0.
\ena
Then from (\ref{deltapk}) and the above, we get $\{\Delta \widehat{ {P}}(k)$, $k=0,1,...\}$ converges, which means that $\lim_{k\to\infty}$$ {1}_N^T$$\Delta $$\widehat{ {P}}(k)$ exists.
From (\ref{aladhhhh}), we have
\begin{equation*}
 {1}_N^T {\lambda}(k+1)= {1}_N^T {\lambda}(0)+\sum_{j=0}^k\sigma(j) {1}_N^T\Delta \widehat{ {P}}(j),\ k=0,1,2...,
\end{equation*}
This together with the convergence of $\{\lambda_i(k),k=0,1,...\}$, $i=1,2,...,N$ leads to that the series $\sum_{j=0}^k\sigma(j) {1}_N^T\Delta \widehat{ {P}}(j)$ converges.

Now we prove that $\lim_{k\to\infty} {1}_N^T \Delta \widehat{ {P}}(k)=0$. We use reduction to absurdity.

Assume that $\lim_{k\to\infty} {1}_N^T\Delta \widehat{ {P}}(k)>0$. Then there is a positive integer $k_0$, and a constant $\omega >0$ such that $ {1}_N^T\Delta \widehat{ {P}}(k)\geq\omega$, $k=k_0$, $k_0+1$, $...$ From Assumption \ref{A7}, we have
\begin{equation*}
\sum_{j=k_0}^k\sigma(j) {1}_N^T\Delta \widehat{ {P}}(j)\geq \omega\sum_{j=k_0}^k\sigma(j)\to\infty,\ k\to\infty,
\end{equation*}
This is in contradiction with the convergence of $\sum_{j=0}^k\sigma(j) {1}_N^T\Delta \widehat{ {P}}(j)$. Thus, $\lim_{k\to\infty} {1}_N^T\Delta \widehat{ {P}}(k)\leq0$. Similarly, one can prove that $\lim_{k\to\infty} {1}_N^T\Delta \widehat{ {P}}(k)\geq0$. Therefore,
\begin{equation*}
\lim_{k\to\infty} {1}_N^T\Delta \widehat{ {P}}(k)=0.
\end{equation*}
Then by (\ref{ppbaaddaddadd1}), we have
\begin{equation*}
\lim_{k\to\infty} {1}_N^T\Delta {P}(k)=\lim_{k\to\infty} {1}_N^T\Delta \widehat{ {P}}(k)=0,
\end{equation*}
which together with (\ref{deltapk}) and (\ref{deltapconveraddadd1}) gives
\begin{equation*}
\lim_{k\to\infty}\Delta \widehat{ {P}}(k)=\lim_{k\to\infty} {J}_N\Delta {P}(k)=\lim_{k\to\infty}\frac{1}{N} {1}_N {1}_N^T\Delta {P}(k)= {0},
\end{equation*}
that is, $\lim_{k\rightarrow\infty}\Delta \widehat{P}_i(k)=0$, $i=1,2,...,N$. Notice that (\ref{ppbaaddaddadd1}) means
\begin{equation*}
\sum_{i=1}^N\Delta \widehat{P}_i(k)+\sum_{i=1}^NP_i(k)=\sum_{i=1}^N P_{Di}+\sum_{i=1}^NB_iP^2_i(k),\ k=0,1,2....
\end{equation*}
This together with $\lim_{k\rightarrow\infty}\Delta \widehat{P}_i(k)=0$, $i=1,2,...,N$ leads to that the balance of supply and demand is  achieved for the microgrid system asymptotically. From the convergence of $\{\lambda_i(k), k=0,1,...\}$, $i=1, 2,..., N$, Theorem \ref{oaddfff} and the fact that for all DGs, the incremental costs with penalty factor are equal and the power supply and demand are balanced for the optimal solution of ED, we have (\ref{optimalisolatedaddadd1}).
\hfill $\blacksquare$

\vskip 0.2cm
Theorems \ref{oaddhhh} and \ref{oaddfff} rely on the equality ${1}_N^T[\Delta \widehat{ {P}}(0)-\Delta {P}(0)]+P_{MG}(0)=0$, that is, $\sum_i^N\Delta \widehat{P}_i(0)=\sum_{i=1}^N P_{Di}+\sum_{i=1}^NP_{Li}(0)-\left(\sum_i^NP_i(0)+P_{MG}(0)\right)$, which means the estimate for the total power mismatch of the microgrid system  is equal to the real total power mismatch of the microgrid system at initial time. This can  be ensured by properly selecting $\Delta \widehat{P}_i(0)$, $i=1,2,...,N$. Then does the estimate still match the real value if mode switching happens at some unpredictable time ? In this case, the estimates by ICUs for the average power mismatch of the microgrid system at the switching moment are not free choices. Fortunately,
we can show that the algorithm (\ref{itlambda})-(\ref{pm}) ensures that the estimate for the total power mismatch of the microgrid system is always equal to the real total power mismatch of the microgrid system no matter the microgrid is in grid-connected or isolated mode and no matter when mode switching happens. The microgrid can perform reliable transition between the grid-connected and isolated operation modes and the ICUs who are not neighbors of the ER do not need to know when mode switching happens.

\vskip 0.2cm

\begin{theorem}   
\label{oadd1}
For the algorithm (\ref{itlambda})-(\ref{pm}), suppose that Assumptions \ref{A4}-\ref{A7} hold. Then
the microgrid system can achieve reliable transformation between isolated and grid-connected modes. That is,
the estimate for the total power mismatch of the microgrid system $\sum_{i=1}^N\Delta \widehat{P}_i(k)$ is always equal to the real total power mismatch of the microgrid system $\sum_{i=1}^N P_{Di}+\sum_{i=1}^NP_{Li}(k)-(\sum_{i=1}^NP_i(k)+P_{MG}(k))$ no matter the microgrid is in grid-connected or isolated mode and no matter when mode switching happens.
\end{theorem}
\vskip 0.2cm

\emph{Proof}:
Without loss of generality, assume that when $k=T+1$, $g$ changes from 1 to 0, that is, the microgrid transits from the grid-connected operation mode to the isolated operation mode.

When $0\leq k\leq T$, the microgrid is in grid-connected mode. From (\ref{deltapppbingwang}), we have
\bna
\label{gridcaddadd1}
 {1}_N^T[\Delta \widehat{ {P}}(k)-\Delta {P}(k)]+P_{MG}(k)=0,\ k=0,1,2,...,T,
\ena
which means
\bna
\label{shouheng1}
 \sum_{i=1}^N\Delta \widehat{P}_i(k)=\sum_{i=1}^N P_{Di}+\sum_{i=1}^NP_{Li}(k)-\left(\sum_{i=1}^NP_i(k)+P_{MG}(k)\right), k=1,...,T.
\ena
That is, the algorithm (\ref{itlambda})-(\ref{pm}) ensures that the estimates for the total power mismatch of the microgrid system ${1}_N^T\Delta \widehat{ {P}}(k)$ is always equal to the real total power mismatch of the microgrid system $ {1}_N^T\Delta {P}(k)-P_{MG}(k)$.

When $k=T+1,T+2,...$, the microgrid is in isolated mode and $P_{Mi}(k)=P_{MG}(k)\equiv0$, $k=T+1$, $T+2$,... From (\ref{itpm235}), it is known that  $\Delta \widehat{ {P}}(T+1)$ depends on  $ {P_M}(T)$ which is not zero as $ {P_M}(0)$.  Next we divide the time interval $k>T $ into $ k = T + 1 $ and $ k> T + 1 $.

When $k=T+1$, it is obtained from (\ref{itpi})-(\ref{itpm235}) and $P_{Mi}(T+1)=0$ that
\begin{eqnarray}
\label{fangcheng22}
\left\{                  
\begin{array}{lll}     
     {P}(T+1)=\Phi( {\lambda}(T+1)),
    \\
    \begin{aligned}
   \Delta \widehat{ {P}}(T+1)&=( {I_N}-\mu^\prime {L_{\mathcal{G}}})\Delta \widehat{ {P}}(T)+\Delta {P}(T+1)-\Delta {P}(T)+ {\mathcal{A}_{*0}} {P_M}(T).
   \end{aligned}
\end{array}
\right.             
\end{eqnarray}
From (\ref{itpm23}) and $P_{Mi}(0)=0$, we get
\ban
&&P_{Mi}(k+1)\cr
&=&P_{Mi}(k)+a_{i0}\Delta P_{Mi}(k+1)\cr
&=&P_{Mi}(k-1)+a_{i0}\Delta P_{Mi}(k)+a_{i0}\Delta P_{Mi}(k+1)\cr
&&...\cr
&=&a_{i0}\sum_{j=1}^{k+1}\Delta P_{Mi}(j),\ k=1,2,...,T,
\ean
which implies
\ban
&&a_{i0}P_{Mi}(k+1)\cr
&=&a_{i0}a_{i0}\sum_{j=1}^{k+1}\Delta P_{Mi}(j)\cr
&=&a_{i0}\sum_{j=1}^{k+1}\Delta P_{Mi}(j)=P_{Mi}(k+1),\ k=1,2,...,T.
\ean
This gives
\begin{equation*}
 {1}_N^T {P_M}(k)= {1}_N^T {\mathcal{A}_{*0}} {P_M}(k),\ k=1,2,...,T.
\end{equation*}
 Then from (\ref{fangcheng22}) and (\ref{gridcaddadd1}), we get
\begin{equation*}
  \begin{split}
  &\quad {1}_N^T[\Delta \widehat{ {P}}(T+1)-\Delta {P}(T+1)]\\
  &= {1}_N^T[\Delta \widehat{ {P}}(T)-\Delta {P}(T)+ {\mathcal{A}_{*0}} {P_M}(T)]\\
  &= {1}_N^T[\Delta \widehat{ {P}}(T)-\Delta {P}(T)+ {P_M}(T)]\\
  &= {1}_N^T[\Delta \widehat{ {P}}(T)-\Delta {P}(T)]+P_{MG}(T)=0.
  \end{split}
  \end{equation*}
which means
\begin{equation}
\label{shouheng2}
 \sum_{i=1}^N\Delta \widehat{P}_i(T+1)=\sum_{i=1}^N P_{Di}+\sum_{i=1}^NP_{Li}(T+1)-\sum_{i=1}^NP_i(T+1).
\end{equation}
 That is, the algorithm (\ref{itlambda})-(\ref{pm}) ensures  that  the estimate for the total power mismatch of the microgrid system $ {1}_N^T\Delta \widehat{ {P}}(k)$  is equal to the real total power mismatch of the microgrid system $ {1}_N^T\Delta {P}(k)$ at the moment when the grid-connected mode is switched to the isolated mode.

 When $k>T+1$, the microgrid is in isolated operation mode, that is, $g=0$. From (\ref{aladhhh2}) and the above equation, we have
 \begin{equation*}
\begin{split}
  &\quad {1}_N^T[\Delta \widehat{ {P}}(k)-\Delta {P}(k)]\\
  &= {1}_N^T[( {I_N}-\mu^\prime {L_{\mathcal{G}}})\Delta \widehat{ {P}}(k-1)+\Delta {P}(k)-\Delta {P}(k-1)-\Delta {P}(k)]\\
  &= {1}_N^T[\Delta \widehat{ {P}}(k-1)-\Delta {P}(k-1)]\\
  &...\\
  &= {1}_N^T[\Delta \widehat{ {P}}(T+1)-\Delta {P}(T+1)]\\
  &=0,\ k=T+2,T+3,...,
\end{split}
\end{equation*}
which means
\begin{equation}
\label{shouheng3}
 \sum_{i=1}^N\Delta \widehat{P}_i(k)=\sum_{i=1}^N P_{Di}+\sum_{i=1}^NP_{Li}(k)-\sum_{i=1}^NP_i(k),\ k=T+2,...,
\end{equation}
that is, the algorithm (\ref{itlambda})-(\ref{pm}) ensures the estimate for the total power mismatch of the microgrid system $ {1}_N^T\Delta \widehat{ {P}}(k)$  is equal to the real total power mismatch of the microgrid system
${1}_N^T\Delta {P}(k)$ in the isolated mode.

Combining (\ref{shouheng1}), (\ref{shouheng2}) and (\ref{shouheng3}), we get
\ban
\sum_{i=1}^N\Delta \widehat{P}_i(k)=\sum_{i=1}^N P_{Di}+\sum_{i=1}^NP_{Li}(k)-\left(\sum_{i=1}^NP_i(k)+P_{MG}(k)\right),\ k=0,1,....
\ean
Similarly, if $g$ changes from 0 to 1 at some time, then the above equality also holds.
\hfill $\blacksquare$

\begin{figure}[!t]
\centering
\begin{minipage}[c]{0.4\textwidth}
\centering
\includegraphics[width=1\textwidth]{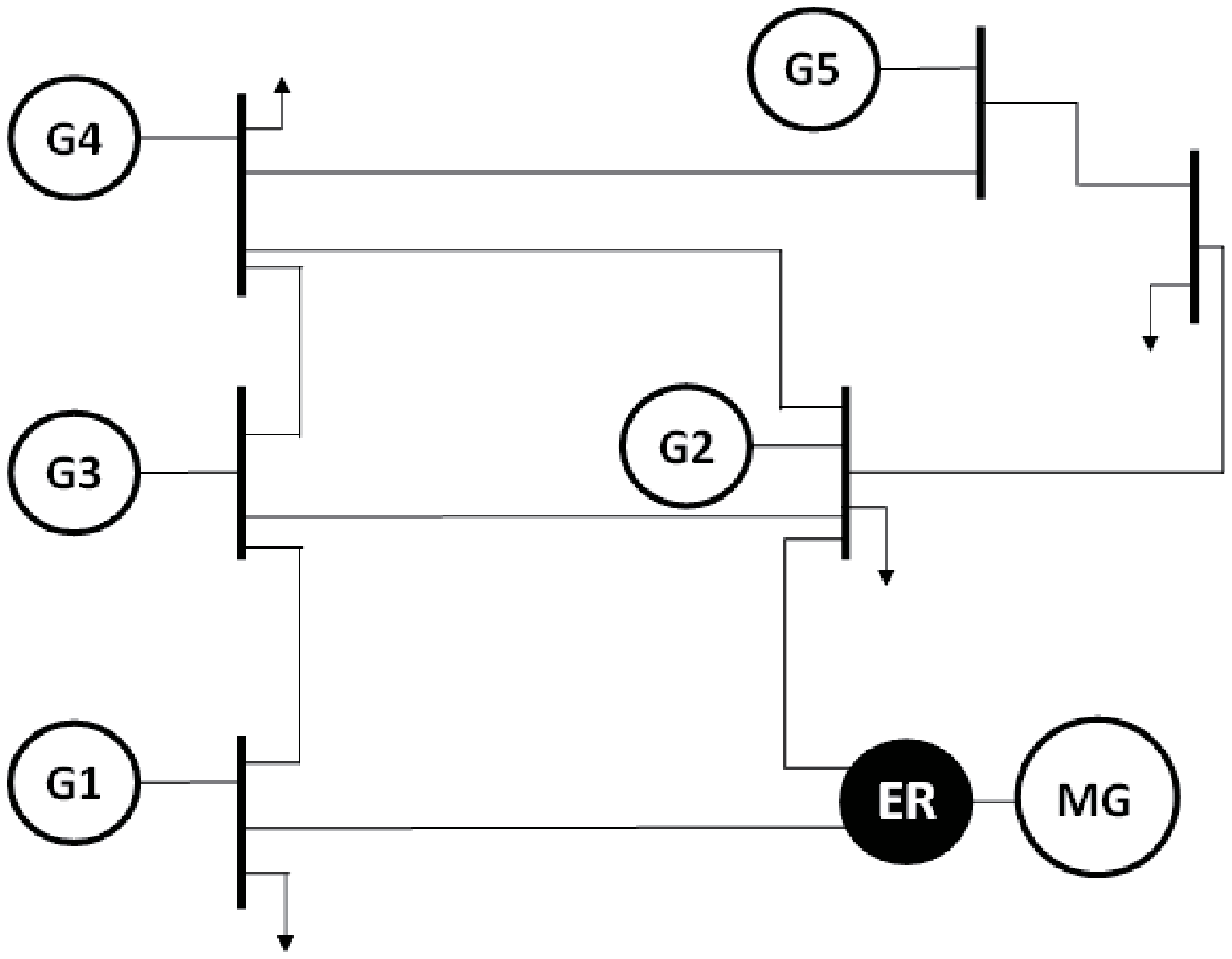}
\end{minipage}
\hspace{0.02\textwidth}
\begin{minipage}[c]{0.4\textwidth}
\centering
\includegraphics[width=1\textwidth]{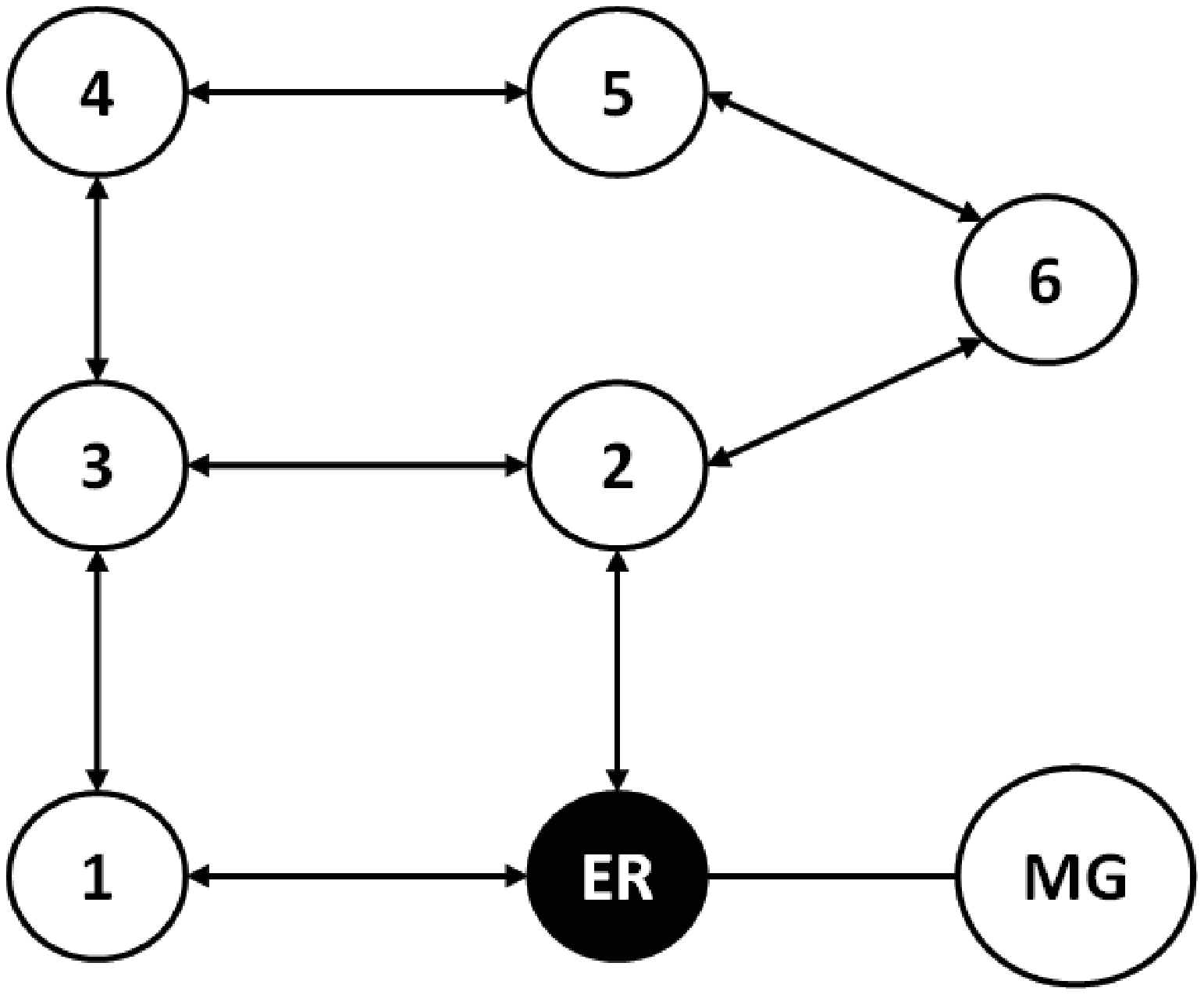}
\end{minipage}\\[3mm]
\begin{minipage}[t]{0.4\textwidth}
\centering
\caption{A test system of Energy Internet.}
\label{fig3}
\end{minipage}
\hspace{0.02\textwidth}
\begin{minipage}[t]{0.4\textwidth}
\centering
\caption{Communication topology of the test system.}
\label{fig4}
\end{minipage}
\end{figure}

\section{NUMERICAL RESULTS}
In this section, we provide two examples to demonstrate
the effectiveness of the proposed algorithms. The electrical network structure of the test system  and the communication network structure among ICUs are shown in Fig. \ref{fig3} and Fig. \ref{fig4}, respectively, containing $5$ DGs, $1$
ER, and $4$ loads. Note that here, the communication network and the electrical network do not share the same structure.
In Fig. 5, Bus 2 and Bus 4 are neighbors in the electrical network, but ICU 2  at Bus 2 and ICU 4 at Bus 4 are not neighbors in the communication network as shown in Fig. 6.
The microgrid is connected to the distribution system through the ER. The parameters of each DG are given
in Table~\ref{tab1}.

\begin{table}[!t]
\renewcommand{\arraystretch}{1.3}
\caption{Parameters of the distributed generations}
\label{tab1}
\centering
\begin{tabular}{|c||c||c||c||c||c||c|}
\hline
DG & $\alpha_i$ & $\beta_i$ & $\gamma_i$  & $\underline{P}_{i}$ & $\overline{P}_i$ &$B_i$ \\
\hline
 G1 & -7830.11 & 93.81 & -326572  & 50 &200 &0.00021\\
\hline
G2 & -4658.77 & 56.24 & -192750    & 20 &70 &0.00017\\
\hline
G3 & -5337.61 & 64.52 & -220578    & 0 &100 &0.00016\\
\hline
G4 & -6047.20 & 73.75 & -247705    & 0 &150 &0.00020\\
\hline
G5 & -5468.96 & 67.48 & -221390    & 45 &180 &0.00019\\
\hline
\end{tabular}
\end{table}

\subsection{Feasibility of grid-connected mode}
 For this case, the total demand of the 4 loads is 550 MW, and the loads at buses 1, 2, 4, and 6 are 50 MW, 150 MW, 150 MW, and 200 MW, respectively. The electricity price of the distribution system obtained by the ER is $85$\yen/MW. The optimal ED solution is given by $P_1^*= 50.000$MW, $P_2^*=46.329$MW, $P_3^*=53.210$MW, $P_4^*=63.165$MW, $P_5^*=83.922$MW and $P_{MG}^*=256.853$MW, which means that the microgrid needs the distribution system to supply power for achieving the optimal ED.

\begin{figure}[!t]
\centering
\includegraphics[width=4.5in]{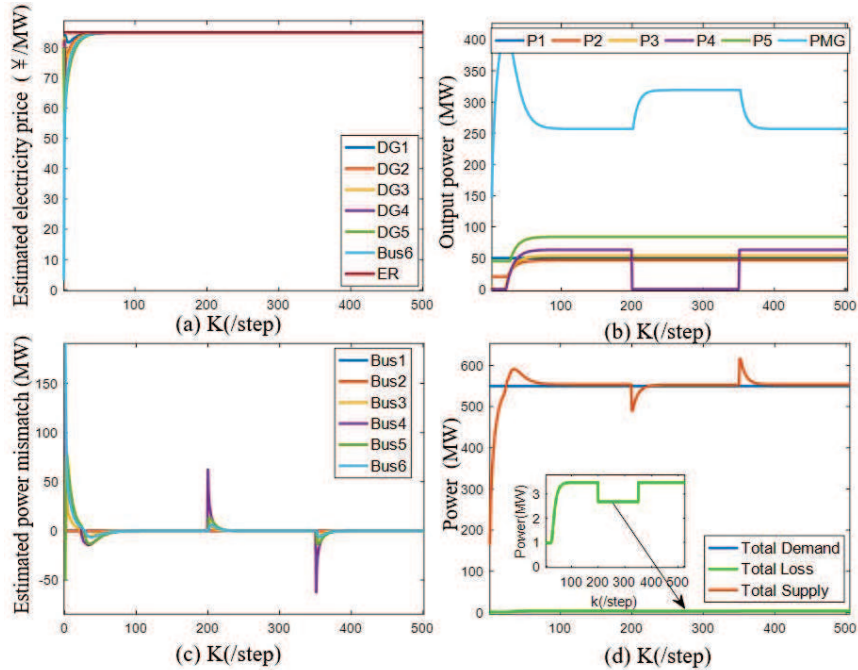}
\caption{Grid connected operation: $\epsilon_i=0.1, i=1,2,...,6; \mu=0.1$ (a) Incremental costs with penalty factors; (b) Active power generated by each DG and active power exchanged with the distribution system; (c) Estimates of ICUs for average power mismatch; (d) Total supply, demand and loss.}
\label{fig5}
\end{figure}

 For the algorithm (\ref{itlambda1})-(\ref{pmguji}) in Section III, the simulation results are shown in Fig. \ref{fig5} when $\epsilon_i=0.1, i=1,2,...,6; \mu=0.1$. We can see that for all DGs, the incremental costs with penalty factor asymptotically converge to the electricity price of the distribution system obtained by the ER exponentially fast. Furthermore, the active power $P_i(k)$ generated by the $i$th DG converges to $P_i^*$, $i=1,2,...,5$, respectively, exponentially fast. The estimated total loss achieves $3.479$MW, and the active power supplied by the power distribution system $P_{MG}(k)$ converges to $P_{MG}^*$.  Further, we demonstrate the effectiveness of the algorithm for the ``plug-and-play'' feature of DGs. At
$k = 200$,  DG $4$ breaks down due to no wind or cloudy weather and at $k = 350$,
DG $4$ reconnects to the microgrid. It can be seen that
the active power generated by each DG and that exchanged with the distribution system response well to status changes.

\begin{figure}[!t]
\centering
\includegraphics[width=4.5in]{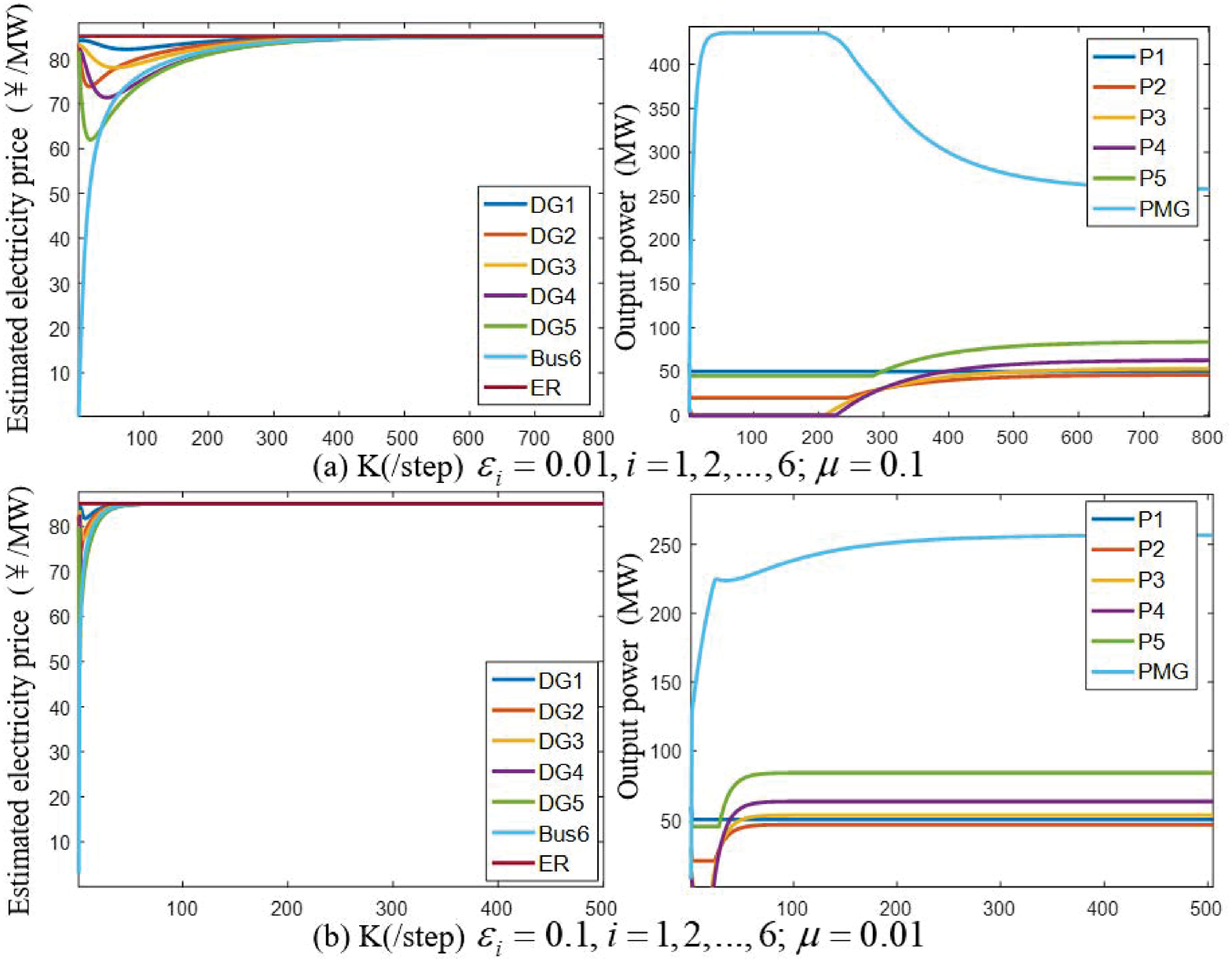}
\caption{Grid connected operation. (a) $\epsilon_i=0.01, i=1,2,...,6; \mu=0.1$; (b) $\epsilon_i=0.1, i=1,2,...,6; \mu=0.01$; }
\label{fig5_1}
\end{figure}

Next, we investigate how the algorithm gains affect the convergence rate.
When $\epsilon_i=0.01$, $i=1,2,...,6$; $\mu=0.1$ and $\epsilon_i=0.1$, $i=1,2,...,6$; $\mu=0.01$ , the simulation results are shown in Fig. \ref{fig5_1}. (a) and (b), respectively. It can be found that if $\epsilon_i$ becomes smaller, then the convergences of incremental cost with penalty factor, the active power generated by each DG and the active power exchanged with the distribution system all become slower. This is mainly due to that the convergences of active power generated by each DG and active power exchanged with the distribution system both depend on  the convergence  of the incremental cost with penalty factor. And when $\mu$ becomes smaller, it only slows down the convergence of the active power generated by each DG and that exchanged with the distribution system.

\subsection{Feasibility of smooth transition between isolated mode and grid-connected mode}

\begin{figure}[!t]
\centering
\includegraphics[width=4.5in]{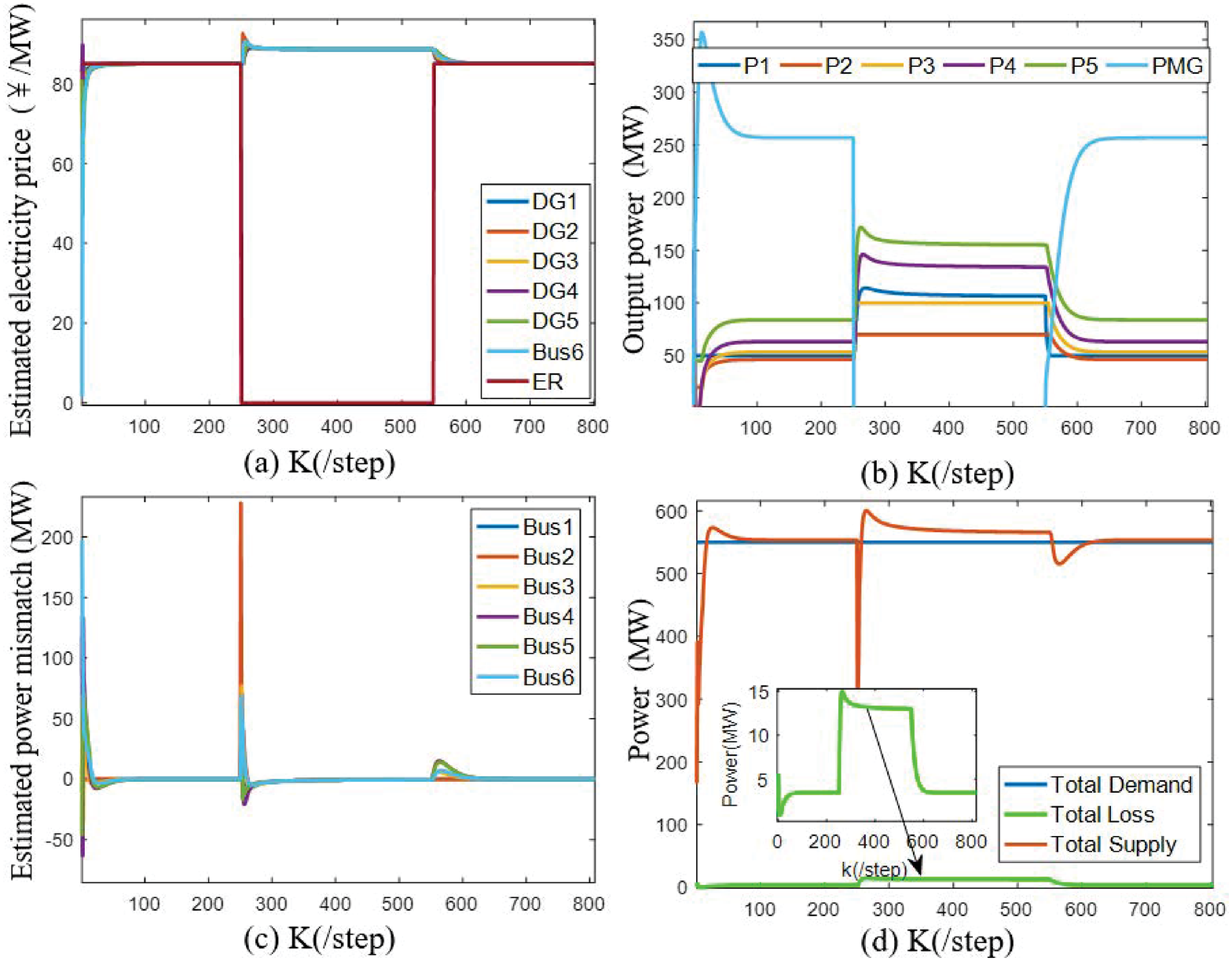}
\caption{Isolated and grid-connected operation: $\epsilon_i^\prime=0.1, i=1,2,...,6; \mu^\prime=0.1; \sigma(k)=\frac{1}{1+k}$ (a) Incremental costs with penalty factors; (b) Active power generated by  each DG and active power exchanged with the distribution system; (c) Estimates of ICUs for average power mismatch; (d)  Total power supply, demand and loss.}
\label{fig6}
\end{figure}

\begin{figure}[!t]
\centering
\includegraphics[width=4.5in]{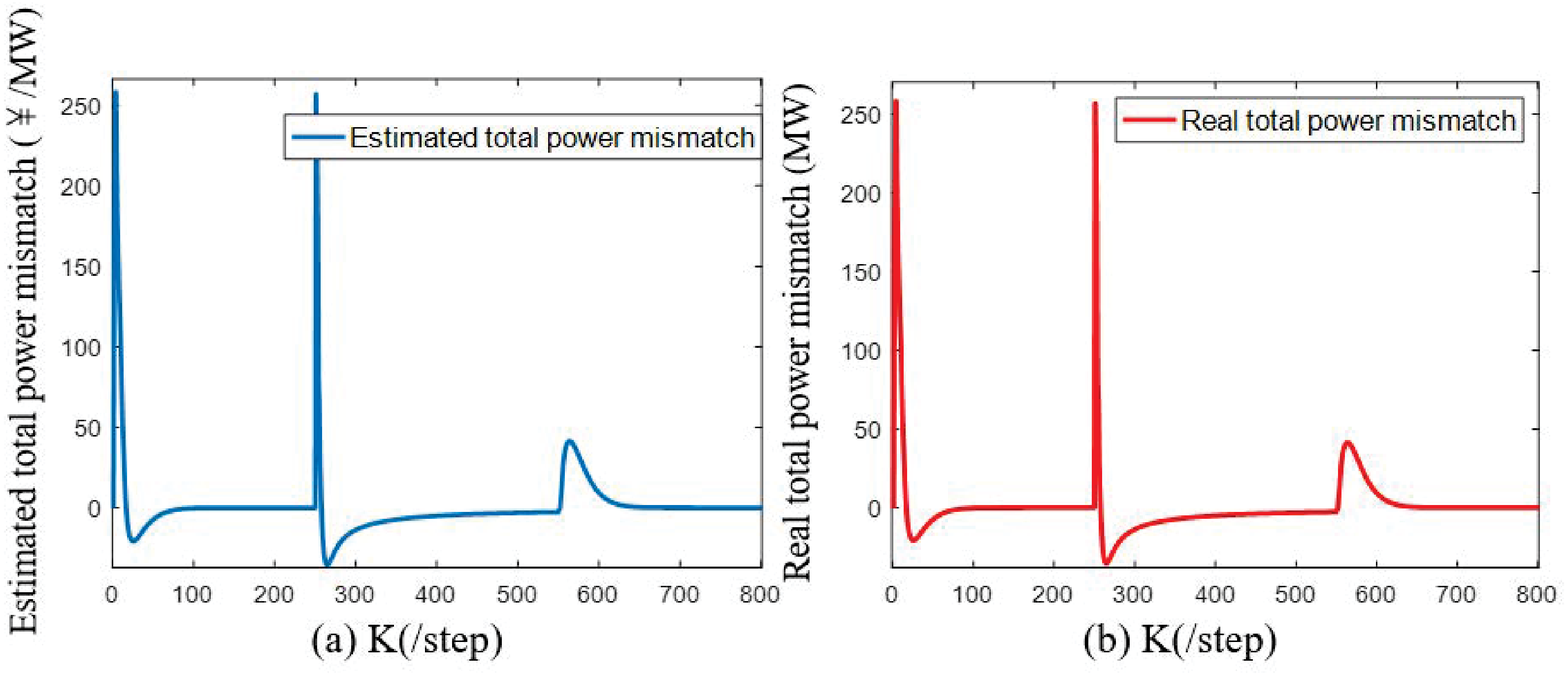}
\caption{Isolated and grid-connected operation: $\epsilon_i^\prime=0.1, i=1,2,...,6; \mu^\prime=0.1; \sigma(k)=\frac{1}{1+k}$. (a) Estimates of total power mismatch; (b) Real total power mismatch.}
\label{fig7}
\end{figure}

This subsection is focused on  the performance of the proposed algorithm  (\ref{itlambda})-(\ref{pm}) in Section IV covering both isolated and grid-connected modes. At
$k = 250$, the distribution system fails and the ER sets the operation mode decision variable $g$ to 0, indicating that the microgrid is switched to isolated mode. At $k = 550$,
the distribution system recovers to normal and the ER sets the operation mode decision variable $g$ to 1, indicating that the microgrid is switched back to grid-connected mode. The optimal  ED solution in isolated mode is given by $P_1^{*\prime}=105.523$MW, $P_2^{*\prime}=70.000$MW, $P_3^{*\prime}=100.000$MW, $P_4^{*\prime}=133.148$MW and $P_5^{*\prime}=154.162$MW. The simulation results are shown in Figures \ref{fig6} and \ref{fig7} with $\epsilon_i^\prime=0.1, i=1,2,...,6; \mu^\prime=0.1$ and
$\sigma(k)=\frac{1}{1+k}$. When the microgrid is switched to isolated mode, the power supplied by the distribution system is cut off immediately. For all DGs, the incremental costs with penalty factor  shown in Fig. \ref{fig6}.(a) converge to the new optimal state $\lambda^{*\prime}=88.541$ \yen/MW; Fig. \ref{fig6}.(b) shows that the active power  $P_i(k)$ generated by the $i$th DGs converges to $P_i^{*\prime}$, $i=1,2,...,5$, respectively. The estimated total loss becomes $12.833$MW. And when the microgrid is switched back to grid-connected mode, the power supplied by the distribution system is recovered, the active power $P_i(k)$ generated by the $i$th DG converges to $P_i^*$ and
the active power supplied by the power distribution system $P_{MG}(k)$ converges to $P_{MG}^*$ once more.
It is shown that the  algorithm
(\ref{itlambda})-(\ref{pm}) converges slower than the algorithm (\ref{itlambda1})-(\ref{pmguji}) due to the vanishing feedback gain $\sigma(k)$.

Fig. \ref{fig7} shows that the estimate for the total power mismatch of the microgrid system is always equal to the real total power
mismatch of the microgrid system no matter the microgrid is in grid-connected or isolated mode and no matter when mode switching happens. The simulation results show that the
microgrid can perform reliable transition between the grid-connected and isolated operation modes.

\section{Conclusion}
 In this paper, distributed ED algorithms for an Energy Internet based on multi-agent consensus control and incremental power exchanged by the ER have been proposed. Firstly, the grid-connected case is considered and all ICUs know that the microgrid is in the grid-connected mode. It is proved that if the communication topology of the Energy Internet contains a spanning tree with the ER as the root node, all ICUs of the microgrid form an undirected graph, and there is a path from each ICU to the ER, then the algorithm can ensure that for all DGs, the incremental costs with penalty factors converge to the electricity price of the distribution system,  the balance of power supply and demand of the whole microgrid is achieved and the optimal ED is achieved asymptotically. Based on the grid-connected algorithm, a fully distributed and autonomous ED algorithm is further proposed which can ensure the smooth switching between the grid-connected and isolated operation modes.  The ICUs which are not neighbors of the ER do not need to know the operation mode of the microgrid. It is proved that if the communication topology of the Energy Internet contains a spanning tree with the ER as the root, the communication topology of the microgrid is connected and there is at least one ICU neighboring the ER bidirectionally, then the algorithm can ensure that the microgrid can reliably transit between the isolated and the grid-connected modes. Finally, the effectiveness of the algorithms is demonstrated by numerical simulations.

The optimal EDP considered in this paper, which is focused on the optimal allocation of active power with the constraints of the balance of power supply and demand  as well as the power generation limits, is a special case of optimal power flow problems.  In an optimal power flow problem, it is necessary to further consider the constraints of various electrical parameters, such as power flow constraints \cite{Dall2013Distributed}, constraints on voltage phase, voltage amplitude, reactive power, frequency \cite{Shi2015Distributed} and line flow constraint \cite{Ma2015Fully}, etc. Then, every bus nodes need to be divided into PV, PQ and balanced nodes.
How to realize the optimal dispatch of the active power of each DG for an Energy Internet in a distributed way with the constraints of the power flow and various electrical parameters would be a challenging issue. One possible idea is to embed distributed line power flow calculations in the optimal power flow algorithm and to estimate the total transmission loss of the system. Since the whole algorithm embeds the algorithms for power flow calculation and estimation of total transmission loss, the convergence condition, convergence precision and rate are all affected by the embedded algorithms. The analysis of the convergence of the whole algorithm  requires a completely different theoretical framework and would be an interesting research topic in future.

Another deficiency of this paper is the usage of the assumption that for all DGs, the incremental costs with penalty factors converge in the convergence analysis of the second proposed algorithm which integrates both the grid-connected and isolated operation modes. Though lots of numerical simulations demonstrate the convergence of the incremental costs with penalty factors to a common value for this algorithm, how to remove this assumption needs far more rigorous
analysis and still remains open. Also, we only  consider optimal ED algorithms on the dispatch level. It is worth studying how to design the corresponding controller  to implement the optimal ED solution on the physical layer. Also, this paper is focused on the case with ideal communication. However, in actual communication networks among ICUs, there must be many uncertainties such as noises, packet dropouts and random switching of communication topologies, which also need future investigation.

Besides the active theoretical research,
at present, several experimental projects for Energy Internet have been in progress, such as the Digital Grid Plan of Japan, which uses Internet technology to carry out experiments in Kenya (\cite{Boyd2013internet}), the``E-Energy'' program in Germany, which sets up six pilot areas in 2008 with thousands of families and hundreds of companies participating in (\cite{Vermesan2011Internet}) and the Future Renewable Electric Energy Delivery and Management System (FREEDM) launched in the USA (\cite{Huang2011future}). It can be expected that more and more challenging theoretical issues  will arise for the control and optimization of Energy Internet in future.
\end{document}